\documentclass[a4paper,11pt]{article}
\usepackage{jheppub} 
\usepackage{lineno}
\usepackage[utf8]{inputenc}

\usepackage{comment}
\usepackage{graphicx}
\usepackage{graphics}
\usepackage{hyperref}
\usepackage{amssymb}
\usepackage{amsmath}
\usepackage{slashed}
\usepackage{multirow}
\usepackage{float}
\usepackage{array}

\usepackage{subfigure}
\usepackage[usenames,dvipsnames]{color} 

\usepackage{graphicx}
\usepackage{dcolumn}
\usepackage{bm}

\arxivnumber{} 

\title{\boldmath Muon collider probes of Majorana neutrino dipole moments and masses}

\author[1]{Michele Frigerio}
\author[2]{Natascia Vignaroli}
\affiliation[1]{Laboratoire Charles Coulomb (L2C), University of Montpellier, CNRS, Montpellier, France}
\affiliation[2]{Dipartimento di Matematica e Fisica, Universit\`{a} del Salento, and 
Istituto Nazionale di Fisica Nucleare, Sezione di Lecce, I-73100 Lecce, Italy}

\emailAdd{\\michele.frigerio@umontpellier.fr, natascia.vignaroli@unisalento.it}

\abstract{ Majorana neutrinos may have transitional dipole moments, which violate lepton number as well as lepton flavour. We estimate the sensitivity of future colliders to the electron-muon neutrino dipole moment, $\lambda_{e\mu}$, by considering same-sign dilepton final states. We find that hadron colliders, even the proposed FCC-hh, are sensitive only to $|\lambda_{e\mu}|\gtrsim 10^{-9}\mu_B$ (with $\mu_B$ the Bohr magneton), a value two-three orders of magnitude larger than current bounds from astrophysics and low-energy neutrino-scattering experiments. In the case of a future muon collider, we show that the sensitivity varies from $|\lambda_{e\mu}|\sim 5\cdot 10^{-9}\mu_B$ for energy $\sqrt{s}\simeq 3$ TeV, to $\sim 10^{-12}\mu_B$ for $\sqrt{s}\simeq 50$ TeV, matching the current laboratory bounds for $\sqrt{s}\simeq 30$ TeV. 
The singular advantage of the muon collider signal would be a direct, clean identification of lepton number and flavour violation.  
We also show that a  muon collider would improve by orders of magnitude the direct bounds on $m_{e\mu}$ and $m_{\mu\mu}$, two of the entries of the Majorana neutrino mass matrix. These bounds could be as strong as $\sim 50$ keV, still far above the neutrino mass scale. 
}

\date{\today}

\begin{document}
\maketitle
\flushbottom

\section{Introduction}

 Neutrino dipole moments (NDMs) are a window into new physics beyond the Standard Model (BSM) \cite{Giunti:2014ixa}.  
Assuming that the neutrino mass eigenstates are Majorana fermions,
only transition dipole moments (TDMs), connecting different flavours, are non-vanishing, and they violate lepton number by two units. 
In the flavour basis, we can denote them by $\lambda_{\alpha\beta}$ with $\alpha,\beta=e,\mu,\tau$, and $\alpha\ne\beta$ since the matrix $\lambda$ is antisymmetric.
Majorana masses induce a one-loop contribution to TDMs, which is suppressed not only by the smallness of neutrino masses, but also by a Glashow-Iliopoulos-Maiani (GIM) mechanism. 
However, the magnitude of the TDMs can significantly increase in new physics models \cite{Voloshin:1987qy, Barr:1990um, Babu:1990hu}. In particular, compelling BSM scenarios  predict a $\nu_\mu$-$\nu_e$ TDM of the order of $\lambda_{e\mu} \sim 10^{-12}\mu_B$, where $\mu_B\equiv e/(2m_e)$ denotes the Bohr magneton (see for example  \cite{Lindner:2017uvt, Zhang:2024ijy} and reference therein). Such large Majorana-neutrino TDMs can be also natural \cite{Bell:2006wi, Davidson:2005cs}, in the light of effective-field-theory (EFT) arguments.
In this range, TDMs could be accessible experimentally, as detailed below.
The observation of lepton number violation (LNV), in a process induced by neutrino TDMs, would additionally establish the Majorana nature of neutrinos.

In the case of Dirac neutrino mass eigenstates, both flavour diagonal and off-diagonal NDMs are possible. The neutrino-mass contribution to the diagonal ones is not GIM-suppressed, but it is still very small, $\lambda_\nu \sim 10^{-19}\mu_B$ \cite{Fujikawa:1980yx}. Larger Dirac NDMs can be generated by new physics, however
values exceeding about $10^{-15}\mu_B$ would not be natural, because they would induce a one-loop contribution to neutrino masses which is unacceptably large \cite{Bell:2005kz}.
In addition, since Dirac NDMs respect lepton number, they do not induce LNV processes at colliders.
We will neglect the Dirac case in the following.

Extensive searches for NDMs have been undertaken in laboratories, as well as through astrophysical and cosmological observations. Laboratory searches typically focus on measuring low-energy scattering of solar neutrinos \cite{Canas:2015yoa, Montanino:2008hu, Borexino:2017fbd, Huang:2018nxj, Coloma:2022umy, Schwemberger:2022fjl, Miranda:2021kre}. 
Astrophysical bounds are obtained from stellar energy loss measurements. While these latter are subject to uncertainties in astrophysical models, they have historically been more stringent than laboratory constraints. 
However, the latest laboratory bounds, $\lambda_\nu < 6.3 \times 10^{-12} \mu_B$ at 90\% C.L. from XENONnT \cite{XENON:2022ltv} and $\lambda_\nu < 6.2 \times 10^{-12} \mu_B$ at 90\% C.L. from LUX-ZEPLIN (LZ) \cite{LZ:2022lsv}, are approaching the astrophysical limits \cite{A:2022acy}. More precisely, these bounds apply to an effective dipole moment \cite{Giunti:2014ixa}, given by 
$\lambda^2_{\nu_e} \simeq\sum_k |U_{ek}|^2\sum_j |\lambda_{jk}|^2$, where $U_{ek}$ is a neutrino mixing-matrix element, and $\lambda_{jk}$ are the neutrino TDMs in the basis of neutrino mass eigenstates, $\lambda_{jk}\equiv \sum_{\alpha\beta} U_{\alpha j}U_{\beta k}\lambda_{\alpha\beta}$.\footnote{ For $E_\nu\gtrsim 1$ MeV, the expression for $\lambda_{\nu_e}$ should be corrected, to account for solar matter effects, which are $E_\nu$-dependent \cite{Giunti:2014ixa}.} Therefore, these searches are sensitive only to a combination of the different TDMs.  
Bounds on $\lambda_{\nu}$  are also placed by short-baseline accelerator and reactor experiments like CE$\nu$NS. These, however, are weaker by at least one order of magnitude, see \cite{Akhmedov:2022txm} and references therein. 
The current most stringent astrophysical limit, $\lambda_\nu \lesssim 10^{-12} \, \mu_B$, is obtained from plasmon decays in the red giant branch of globular 
clusters \cite{Capozzi:2020cbu}.
Cosmological bounds of the order of  $\lambda_\nu \lesssim 3 \times 10^{-12} \mu_B$ can be also obtained from CMB and BBN measurements of the effective number of neutrinos \cite{Li:2022dkc}. 

Another set of constraints is obtained from measurements of neutrino-to-antineutrino conversion in the solar magnetic field \cite{KamLAND:2021gvi}. The sensitivity to $\lambda_\nu$ in this case depends crucially on the value of the solar transverse magnetic field, which is poorly known, so that a conservative bound on $\lambda_\nu$ is typically weaker than those above \cite{Akhmedov:2022txm}. On the other hand, in contrast with all observables considered previously, the observation of an antineutrino flux from the Sun would be a direct evidence for LNV.

In this study we will explore for the first time the sensitivity to NDMs, in particular to the $\nu_\mu$-$\nu_e$ TDM, of collider experiments. These could provide important complementary tests to those that can come from astrophysics and low-energy scattering experiments. We will focus on same-sign dilepton final states, that could provide a direct evidence for LNV, and a clean measurement of $\lambda_{e\mu}$.
Notice that here the challenge is to observe LNV from light neutrino interactions only, to be contrasted with the more common attempt to observe LNV at colliders from new physics at the TeV scale, for example in seesaw models, as reviewed e.g.~in \cite{Cai:2017mow}.

We will consider current and future hadron colliders: the upcoming high-luminosity phase of the LHC (HL-LHC) and the proposed FCC-hh  \cite{FCC:2018vvp}, as well as a possible future multi-TeV muon collider experiment. The latter has been considered among the key points of the strategic plans for the development of particle physics both in Europe \cite{Accettura:2023ked, MuCol-nature} and in the United States \cite{Narain:2022qud, MuCol-nature} and it has developed a considerable community effort to explore its potential for the discovery of BSM physics (see for example \cite{Capdevilla:2024bwt, Barducci:2024kig, Jana:2023ogd, Liu:2023jta, Vignaroli:2023rxr, Bottaro:2021srh, Asadi:2023csb, Spor:2022hhn, Senol:2022snc, Chen:2022msz, Capdevilla:2021kcf, Ruiz:2021tdt, Chiesa:2021qpr, Franceschini:2021aqd, Asadi:2021gah, Bhattacharya:2023beo, Yin:2020afe} for some of the most recent studies). 

The LNV signatures which we consider to probe neutrino TDMs at colliders are also sensitive to Majorana neutrino masses, more precisely to the mass matrix entries in the flavour basis, $m_{\alpha\beta}$, which also violate lepton number by two units. 
Neutrinoless double beta decay ($0\nu\beta\beta$) experiments can reach a very high sensitivity on $|m_{ee}|$, of order $\sim$0.1 eV \cite{Dolinski:2019nrj}. In particular, 90\% C.L. limits of 79-180 meV and 61-165 meV are placed respectively by GERDA \cite{GERDA:2020xhi} and KamLAND-Zen \cite{KamLAND-Zen:2016pfg}. Indirect constraints on the other entries of the Majorana mass matrix can be inferred from this bound on $m_{ee}$, with some little amount of theoretical prejudice. On the other hand, colliders
can be directly sensitive to $m_{\alpha\beta}$ for $\alpha\beta \ne ee$,
providing complementary information on the nature of neutrino masses.

The hadron collider sensitivity on $|m_{\mu\mu}|$ has been recently explored in \cite{Fuks:2020zbm}, which estimates sensitivities ranging from $\sim$ 5.4 GeV at the HL-LHC to $\sim$ 1.2 GeV at the FCC-hh. Experimental searches \cite{ATLAS:2024rzi, ATLAS:2023tkz, CMS:2022hvh} have been also performed at the 13 TeV LHC with 140 fb$^{-1}$, giving the following 95\% C.L. bounds on Majorana masses: $|m_{e\mu}|<$13 GeV, $|m_{ee}|<$24 GeV \cite{ ATLAS:2024rzi} and $|m_{\mu\mu}|<$16.7 GeV \cite{ATLAS:2023tkz}. 
The recent study in \cite{Li:2023lkl} also presents projected sensitivities on Majorana masses of a future same-sign muon collider. The study reports sensitivities similar to those of the FCC-hh for a 30 TeV collision energy. 
The $B-$ and $K-$meson factories like LHCb  or NA62 experiments are also sensitive to $m_{\mu\mu}$ or $m_{e\mu}$ through LNV meson decay processes like $B^\pm(K^\pm) \to \pi^\mp \ell^\pm \ell^\pm$ \cite{LHCb:2018roe, NA62:2019eax}. The bounds derived in \cite{Fuks:2020zbm} on the basis of the study in \cite{Atre:2005eb} show, however, that their sensitivities would be lower than the projected sensitivities of the FCC-hh: NA62 can test with its 2017 data set masses of the order of 50 GeV, while LHCb sensitivities with 300 fb$^{-1}$ are about five orders of magnitude worst \cite{Fuks:2020zbm}. 
Ref.~\cite{Atre:2005eb} also reviews the possibility to use muon-to-positron conversion to set a bound on $m_{e\mu}$, which is however subject to large uncertainties from nuclear matrix elements.

In this work, we will identify for the first time a strategy to probe Majorana masses at a future $\mu^+\mu^-$ muon collider, and we will present the corresponding projected sensitivities on $m_{e\mu}$ and $m_{\mu\mu}$. 

The paper is organized as follows.
We introduce the theoretical framework for neutrino TDMs in section~\ref{sec:theo}. The possible probes of TDMs at hadron colliders are described in section~\ref{sec:hadron-colliders}, while we present in section~\ref{sec:MuCol}
our search for TDMs in the case of a future muon collider. In section~\ref{sec:masses}, we employ the muon collider to test Majorana neutrino masses. We finally discuss our results 
in section~\ref{sec:conclusions}.  

\section{Neutrino dipole moments: theoretical framework}\label{sec:theo}

Majorana neutrinos can have  
transition dipole moments (TDMs), that is to say, dipole moments off-diagonal in flavour space \cite{Schechter:1981hw,Nieves:1981zt,Shrock:1982sc, Kayser:1982br}. They can be described by five-dimensional operators, 
\begin{equation}\label{eq:NMM-A}
\begin{array}{c}
{\cal L} \supset \dfrac{1}{4} \, \overline{(\nu_{L\alpha})^c} \sigma^{\mu\nu} \lambda_{\alpha\beta} \nu_{L\beta} A_{\mu \nu} + \text{h.c.}
= \dfrac 14 \, \overline{\nu_\alpha} \sigma^{\mu\nu} 
(\mu_{\alpha\beta} +i d_{\alpha\beta}\gamma_5)
\nu_\beta A_{\mu \nu}\,,\\ \\
\mu_{\alpha\beta}\equiv i {\rm Im}\lambda_{\alpha\beta} \,,\qquad 
d_{\alpha\beta}\equiv i{\rm Re}\lambda_{\alpha\beta}\,,
\end{array}
\end{equation}
where $\nu_L$ are left-handed spinors, $\nu\equiv \nu_L + (\nu_L)^c $ are Majorana spinors,
$\sigma^{\mu\nu}\equiv (i/2)[\gamma^\mu,\gamma^\nu]$,
$A_{\mu\nu}$ is the photon field strength,
and $\alpha,\beta$ are flavour indexes. 
The TDM matrix  $\lambda$ has mass-dimension minus one, and it is antisymmetric in flavour space, $\lambda_{\alpha\beta}=-\lambda_{\beta\alpha}$. 
The magnetic dipole moments (MDMs) $\mu_{\alpha\beta}$ and the electric dipole moments (EDMs) $d_{\alpha\beta}$ are defined to be imaginary.
The prefactor $1/4$ guarantees the standard normalisation for the $\bar\nu \nu \gamma$ vertex \cite{Grimus:1997aa, AristizabalSierra:2021fuc, deGouvea:2022znk}.
Like the Majorana neutrinos masses $m_{\alpha\beta}$, defined later by Eq.~\eqref{MajM}, the neutrino TDMs $\lambda_{\alpha\beta}$ are associated to operators with lepton number equal to two, therefore a non-zero TDM implies a violation of lepton number by two units.

The electromagnetic dipole operators in Eq.~\eqref{eq:NMM-A} should be the low-energy realisation of $SU(2)_w\times U(1)_Y$ invariant operators, generated by some LNV new physics at scale $\Lambda_{NP}$.
The lowest-dimensional such operators contributing to neutrino TDMs are two dimension-7 operators~\cite{Davidson:2005cs},
\begin{align}
\label{eq:OB} (O_B)_{\alpha\beta} & = g^\prime \left( \overline{\ell_{L\alpha}^{\,c}}\, \epsilon H \right) \sigma^{\mu\nu} \left(H^T \epsilon  \ell_{L\beta} \right) B_{\mu\nu}\,, \\ 
\label{eq:OW} (O_W)_{\alpha\beta} & = i g \varepsilon_{abc} \left( \overline{\ell_{L\alpha}^{\,c}}\, \epsilon \sigma^a \sigma^{\mu\nu} \ell_{L\beta} \right) \left(  H^T \epsilon \sigma^b H  \right) W^c_{\mu\nu} \, ,
\end{align}
where $H$ and $\ell_L$ are the Higgs and lepton $SU(2)_w$ doublets, 
the $\sigma^a$ are the Pauli matrices, $\epsilon=-i \sigma^2$, $W_{\mu\nu}^a$ and $B_{\mu\nu}$ are the $SU(2)_w$ and $U(1)_Y$ gauge field strengths.
In unitary gauge, $H=[0\;(v+h)/\sqrt{2}]^T$. After 
spontaneous symmetry breaking (SSB), setting to zero the Higgs boson field $h$, and going to the basis of physical gauge bosons, one obtains the following effective interactions: 
\begin{align}
\label{eq:OB-interaction} (O_B)_{\alpha\beta}|_v  &= -\dfrac{g^\prime v^2}{2} \left( \overline{\nu^{\,c}_{L\alpha}}\,  \sigma^{\mu\nu} \nu_{L\beta} \right) (c_w A_{\mu\nu}-s_w Z_{\mu\nu})\,, \\ 
\label{eq:OW-interaction} (O_W)_{\alpha\beta}|_v  &= -g v^2 \left( \overline{\nu^{\,c}_{L\alpha}}\,  \sigma^{\mu\nu} \nu_{L\beta} \right) \left(s_w A_{\mu\nu}+c_w Z_{\mu\nu}+2ig W^-_\mu W^+_\nu \right) \nonumber \\
 - \dfrac{g v^2}{\sqrt{2}}&
\left( \overline{\nu^{\,c}_{L\alpha}}\, \sigma^{\mu\nu} e_{L\beta} + \overline{e^{\,c}_{L\alpha}}\, \sigma^{\mu\nu} \nu_{L\beta} \right) \left[W^+_{\mu\nu} 
+2ig W^+_\mu(s_w A_\nu +c_w Z_\nu)
\right]  \,. 
\end{align}

Assuming ${\cal L}\supset C^B_{\alpha\beta} (O_B)_{\alpha\beta} + C^W_{\alpha\beta} (O_W)_{\alpha\beta} + \text{h.c.}$, the (tree-level) matching onto the neutrino TDM reads\footnote{
In order to compare with  \cite{Davidson:2005cs}, note their $\mu_{\alpha\beta}$
corresponds to $\lambda_{\alpha\beta}/2$, and they use the convention $v\simeq 174$ GeV while here we take $v\simeq 246$ GeV.}
\begin{equation}
\lambda_{\alpha\beta}= -2ev^2\left(C^B_{\alpha\beta}+2C^W_{\alpha\beta}\right)~.
\label{match}\end{equation}
The operator $O_W$ will be of particular relevance for our analysis at hadron colliders. 
For this part of the analysis, we will work under the simplifying assumption that new physics above the electroweak scale generates only $O_W$.
For the muon collider analysis, we will study the effect of both $O_B$ and $O_W$.

One should keep in mind that new physics may violate lepton number {\it without} inducing a neutrino TDM, at least at tree level. 
Indeed, if new physics happens to generate $C^B=-2C^W$, the combination in Eq.~\eqref{match} vanishes, still one could observe same-sign different-flavour dileptons at colliders.
More in general, other dimension-7 operators violate lepton number by two units, without contributing to neutrino TDMs. In particular, let us consider the operator
\begin{equation}\label{eq:OWW} 
(O\rotatebox[origin=c]{180}{$_W$})_{\alpha\beta}  = g 
\left( \overline{\ell_{L\alpha}^{\,c}}\, \epsilon \sigma^{\mu\nu} \ell_{L\beta} \right) \left(  H^T \epsilon \sigma^a H  \right) W^a_{\mu\nu} \, .
\end{equation}
It involves exactly the same fields as $O_W$, but with a different contraction of $SU(2)_w$ indexes, such that $O\rotatebox[origin=c]{180}{$_W$}$ is actually symmetric in flavour, 
$(O\rotatebox[origin=c]{180}{$_W$})_{\alpha\beta}=(O\rotatebox[origin=c]{180}{$_W$})_{\beta\alpha}$.\footnote{
We remark that the two operators, $O_W$ and $O\rotatebox[origin=c]{180}{$_W$}$, both involve two lepton doublets, two Higgs doublets and one $SU(2)_w$ field strength, and they are clearly independent. In the classification of dimension-7 operators presented in Refs.~\cite{Lehman:2014jma,Liao:2016hru}, only one operator involving this same set of fields is displayed: 
$$
(O_{LHW})_{\alpha\beta}\equiv \dfrac 12\left( \overline{\ell_{L\alpha}^{\,c}}\, \epsilon H \right)
\sigma^{\mu\nu} \left(H^T \epsilon \sigma^a \ell_{L\beta} \right) W^a_{\mu\nu}
= \dfrac{1}{4g}\left[(O_W)_{\alpha\beta}+(O\rotatebox[origin=c]{180}{$_W$})_{\alpha\beta} \right]\,, 
$$
where the equality can be explicitly checked in components, or using identities among Pauli matrices. Since $O_W$ is flavour antisymmetric and $O\rotatebox[origin=c]{180}{$_W$}$ is flavour symmetric, the 3 independent Wilson coefficients $C^W_{\alpha\beta}$ and the 6 independent $C^{\rotatebox[origin=c]{180}{$_W$}}_{\alpha\beta}$ are in one-to-one correspondence with the 9 $C^{LHW}_{\alpha\beta}$. In other words, the linear combination $O_{LHW}$ covers all independent operators in this class. Nonetheless, since the physics implications of $O_W$ and $O\rotatebox[origin=c]{180}{$_W$}$ are substantially different, we decided it is preferable to treat them separately.}

Once the Higgs doublet is replaced by its vev, one obtains
\begin{equation} 
\label{eq:OWW-interaction} (O\rotatebox[origin=c]{180}{$_W$})_{\alpha\beta}|_v  = 
 \dfrac{g v^2}{\sqrt{2}}
\left(-\overline{\nu^{\,c}_{L\alpha}}\, \sigma^{\mu\nu} e_{L\beta} + \overline{e^{\,c}_{L\alpha}}\, \sigma^{\mu\nu} \nu_{L\beta} \right) \left[W^+_{\mu\nu} 
+2ig W^+_\mu(s_w A_\nu +c_w Z_\nu)
\right]  \,. 
\end{equation}
Such effective interactions can produce same-sign dileptons at colliders, with either equal or different flavours. 
This or other $\Delta L =2$ operators may induce a neutrino TDM via operator mixing, that is, they may contribute to the Wilson coefficient of Eq.~\eqref{match} at loop level.
An exception is the case of lepton-flavour conserving new physics, which may induce e.g.
$(O\rotatebox[origin=c]{180}{$_W$})_{\alpha\alpha}$, but it does not generate neutrino TDMs even at loop level.

Since we are interested in setting the strongest possible constraint on the neutrino TDMs, in the following we will focus on $O_W$ or $O_B$ only.\footnote{
A recent survey of LNV dimension-7 operators has been presented in \cite{Fridell:2023rtr}. In this study, bounds from low-energy experiments and from current and future hadron colliders are derived on the different operators. In the case of collider bounds, however, the analysis does not consider different lepton flavors in the final state, therefore an analysis of the $O_B$ and $O_W$ operators here considered is missing. }

It is interesting to give an estimate of the relevant Wilson coefficients, by using Naive Dimensional Analysis (NDA). Calling $m_*$ and $g_*$ the typical mass and coupling of the heavy particles in the ultraviolet (UV) theory, and recalling that dipole operators can only arise from loops, one can write
\begin{equation} 
C^{B,W}_{\alpha\beta} = c^{B,W}_{\alpha\beta} \dfrac{1}{(4\pi)^2} 
\dfrac{(g_*\epsilon_H)^2 (g_*\epsilon_\alpha)(g_*\epsilon_\beta)}{m_*^3}\,,
\end{equation}
where $\epsilon_H$ ($\epsilon_\alpha$) quantifies the coupling between the SM Higgs doublet $H$ (lepton doublet $\ell_{L\alpha}$) and the UV particles, in units of $g_*$, while the coefficients $c^{B,W}_{\alpha\beta}$ are expected to be of order one. Using Eq.~(\ref{match}), we obtain
\begin{equation} 
\dfrac{|\lambda_{\alpha\beta}|}{\mu_B} \simeq 2\cdot 10^{-11} |c^B_{\alpha\beta}+2 c^W_{\alpha\beta}|
 \left(\dfrac{10\ {\rm TeV}}{m_*}\right)^3 \left(\dfrac{g_*}{4\pi}\right)^4
\left(\dfrac{\epsilon_H}{1}\right)^2 \left(\dfrac{\epsilon_\alpha \epsilon_\beta}{10^{-3}}\right) 
\,,\label{NDA}
\end{equation}
where $g_*=4\pi$ corresponds to a strongly-coupled UV theory, and $\epsilon_H=1$ corresponds to a composite Higgs. The parameters $\epsilon_\alpha$ quantify the degree of partial compositeness of lepton doublets, and they should be sufficiently small to comply with precision lepton observables, especially lepton-flavour-violating ones \cite{Frigerio:2018uwx}.\footnote{
Assuming a `flavour-anarchic' scenario and fixing $m_*=10$ TeV, the reference value of Eq.~(\ref{NDA}), $\epsilon_\alpha\epsilon_\beta =10^{-3}$, is sufficiently small to respect most constraints, except those from $e-\mu$ transitions, which would require $\epsilon_e \epsilon_\mu \lesssim 3\cdot 10^{-6}$  \cite{Frigerio:2018uwx}.
However, one can consider instead `flavour-symmetric' scenarios, where some UV flavour symmetry suppresses dangerous processes, effectively allowing for larger  
$\epsilon_e\epsilon_\mu$ \cite{Frigerio:2018uwx}.}
After deriving the collider bounds on $\lambda_{e\mu}$, we will come back to Eq.~(\ref{NDA}) in section \ref{sec:conclusions}, to discuss the limitations of the EFT approach.\footnote{We thank Julian Heeck and a Referee for raising this interesting question.}

\section{Testing neutrino dipole moments 
  at hadron colliders}\label{sec:hadron-colliders}

\begin{figure}
\centering
 \includegraphics[scale=0.5]{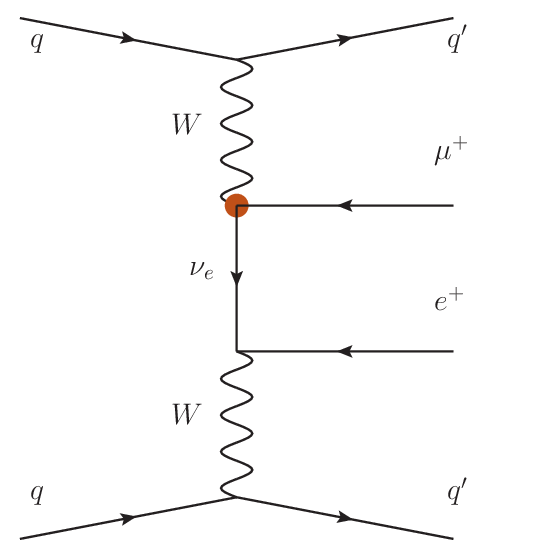} \quad \includegraphics[scale=0.55,trim = 0 -40 0 0]{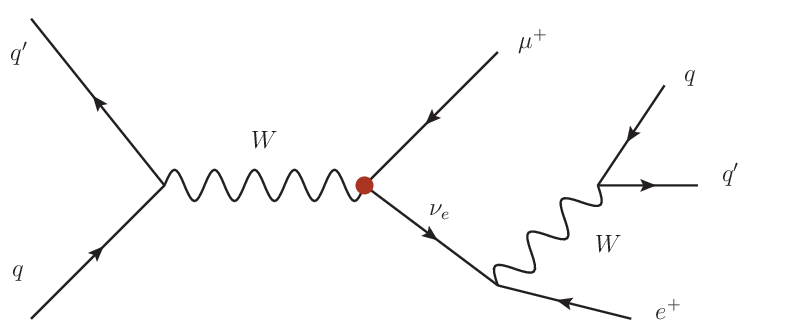}
 \caption{\em Leading Feynman diagrams for the $\Delta L=2$ process induced by an electron-muon neutrino TDM at $pp$ colliders. We also included the analogous diagrams with a muon neutrino exchange, and/or with negative-charge leptons in the final state.
}\label{fig:NMM-pp}
 \end{figure}

We focus on neutrino dipole transitions from the electron flavour to the muon flavour.
At hadron colliders, 
it is possible to probe TDMs generated by the $O_W$ operator, through the effect of the $W$-lepton interactions contained in Eq.~\eqref{eq:OW-interaction},
\begin{equation}\label{eq:NMM-W}
{\cal L}\supset 2\sqrt{2} g C^W_{e\mu} v^2 (\overline{\nu_{L\mu}^{\,c}} \sigma^{\mu\nu} e_L 
- \overline{\nu^{\,c}_{Le}}\sigma^{\mu\nu} \mu_L) \partial_\mu W^+_\nu +h.c.\ ,
\end{equation} 
where we added the two identical contributions from $\alpha\beta=e\mu,\mu e$.
The signal to search for is the $\Delta L=2$ process with two same-sign different-flavour leptons, accompanied by two jets, whose dominant Feynman diagram is shown in Fig. \ref{fig:NMM-pp}.

\begin{figure}
\centering
 \includegraphics[scale=0.7]{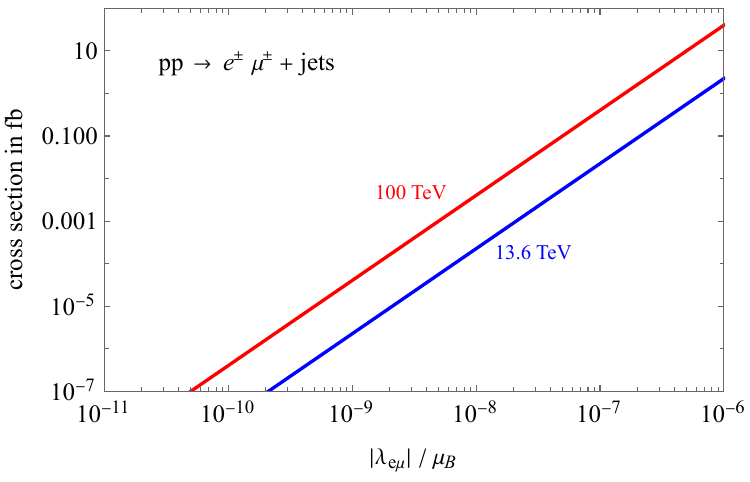}
 \caption{\em  Signal cross section values 
 at a hadron collider with $\sqrt{s}=13.6$ TeV, corresponding to the Run-3-LHC and HL-LHC collision energy, and with $\sqrt{s}=100$ TeV, the nominal value for the FCC-hh. We assumed acceptance cuts as specified in the text.}\label{fig:xsec-pp}
 \end{figure}

 We include the effective $W$ interaction in Eq.~\eqref{eq:NMM-W} in {\tt MadGraph }5 \cite{Alwall:2014hca} by using {\tt Feynrules} \cite{Alloul:2013bka}, and calculate the cross section, at LO in QCD, at a hadron collider with $\sqrt{s}=13.6$ TeV, corresponding to the High-Luminosity LHC collision energy and the current LHC Run-3, and with $\sqrt{s}=100$ TeV, the nominal value for the future FCC-hh collider \cite{FCC:2018vvp}. We apply acceptance cuts of 20 GeV on the $p_T$ of the two leptons and the jets and we require the leptons to lie in the central region, with a rapidity $|\eta|<2.5$, and the jets in the detector region, $|\eta|<5$. The cross section values we obtain, as function of the absolute value of $\lambda_{e\mu}$ (in units of $\mu_B$) are shown in Fig.~\ref{fig:xsec-pp}, where we used the relation in  Eq.~\eqref{match} 
     to relate $C^W_{e\mu}$ to $\lambda_{e\mu}$. We consider both positive and negative charges for the final-state leptons: because of the proton structure, the cross section for the positive-charge configuration is roughly two times larger than the one for negative lepton charges.

The ATLAS collaboration has recently performed a search for heavy Majorana neutrinos in $e^\pm e^\pm$ and $e^\pm \mu^\pm$ final states via $WW$ scattering in $pp$ collisions at $\sqrt{s}$ = 13 TeV \cite{ATLAS:2024rzi}. Since this search focuses on the same final state we are considering, we can recast it to obtain projected sensitivities of hadron colliders to the $\nu_e-\nu_\mu$ TDM.

The dominant background component is given by SM processes producing two same-sign leptons, which are mainly given by $WZ$ and $W^\pm W^\pm$ production in association with two jets. Other, subdominant, sources of background are represented by SM processes with two opposite-sign leptons where the charge of one of the leptons is
misidentified, and by non-prompt lepton production events, where leptons are generated from decays of long-lived particles and mainly arise from semileptonic decays of heavy-flavour hadrons. 

The ATLAS signal selection strategy in the $e^\pm \mu^\pm$ channel requires two jets with invariant mass $m_{jj}>500$ GeV and rapidity separation $|\Delta y_{jj}|>2$. The transverse momentum cuts on the leading-$p_T$ and second-leading-$p_T$ jets, $p_T j(1)>30$ GeV, $p_T j(2)>25$ GeV, are applied. It is also applied a condition on the azimuthal separation of the two leptons, $|\Delta \phi_{e\mu}|>2.0$.
We find that the efficiency of this selection to our neutrino TDM signal is of 0.40 at the HL-LHC and of about 0.70 at the FCC-hh. Note that the selection singles out the $t-$channel signal topology (left diagram in Fig.~\ref{fig:NMM-pp}), while the $s-$channel contribution (diagram on the right) gets suppressed from the kinematic requirements on the jets.
For the background, ATLAS estimates a total SM background of $1.51 \pm 0.08$ fb. 
We use this estimate for HL-LHC.
Based on this value, we can also infer the magnitude of the SM background at the FCC-hh, by considering a scaling with the different collision energy of the cross section for the dominant background processes.  
We evaluate that the background would be of the order of 13 fb at the FCC-hh. This is a rather conservative estimate, since the background to our signal could be reduced with a tailored analysis at future colliders. For example, one could exploit the higher energy of the $jj$ system at the FCC-hh, by applying a stronger cut on $m_{jj}$, or the high statistics collected at the HL-LHC, with the possible application of Boosted-Decision-Tree strategies.

Estimating the statistical significance (i.e.~the number of $\sigma$'s) by the ratio $S/\sqrt{S+B}$, with $S$ ($B$) denoting the signal (background) number of events, we find the following 2$\sigma$ sensitivities: 
\begin{align}
\begin{split}
 \text{HL-LHC} \qquad  &  |\lambda_{e\mu}| < 4.0 \cdot 10^{-7}\, \mu_B \quad \{300\, \text{fb}^{-1}\} , \quad  |\lambda_{e\mu}| < 2.2 \cdot 10^{-7}\, \mu_B \quad \{3\, \text{ab}^{-1}\} ,\\
 \text{FCC-hh} \qquad  &|\lambda_{e\mu}| < 6.8 \cdot 10^{-8}\, \mu_B \quad \{3\, \text{ab}^{-1}\} , \quad  |\lambda_{e\mu}| < 3.8 \cdot 10^{-8}\, \mu_B \quad \{30\, \text{ab}^{-1}\}.
\end{split}
\end{align}
In the most optimistic case in which the background could be reduced to a negligible level, 
the sensitivities would be
\begin{align}
\begin{split}
 \text{HL-LHC} \qquad  & |\lambda_{e\mu}| < 1.2 \cdot 10^{-7}\, \mu_B \quad \{300\, \text{fb}^{-1}\} , \quad |\lambda_{e\mu}| < 3.8 \cdot 10^{-8}\, \mu_B \quad \{3\, \text{ab}^{-1}\},\\
 \text{FCC-hh} \qquad  & |\lambda_{e\mu}| < 6.8 \cdot 10^{-9}\, \mu_B \quad \{3\, \text{ab}^{-1}\} , \quad  |\lambda_{e\mu}| < 2.0 \cdot 10^{-9}\, \mu_B \quad \{30\, \text{ab}^{-1}\}.
\end{split}
\end{align}
We can see that, even for the FCC-hh in the most optimistic scenario of a background reduced to a negligible level, the hadron collider sensitivity would be about three orders of magnitude 
worst than the latest astrophysical and laboratory bounds.
In fact, Eq.~(\ref{NDA}) shows that new physics at scales $m_*\gtrsim \sqrt{s}$ cannot generate $|\lambda_{e\mu}| \gg 10^{-11}\mu_B$, therefore these bounds are also too weak to be interpreted in the EFT  for the neutrino dipole.

\section{Testing neutrino dipole moments  at a muon collider}\label{sec:MuCol}

\begin{figure}
\centering
 \includegraphics[scale=0.55]{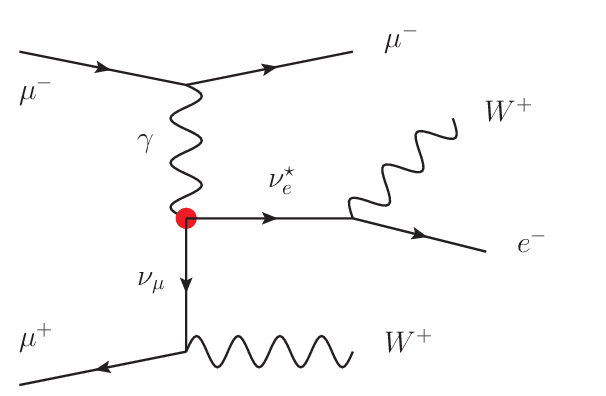} \includegraphics[scale=0.55]{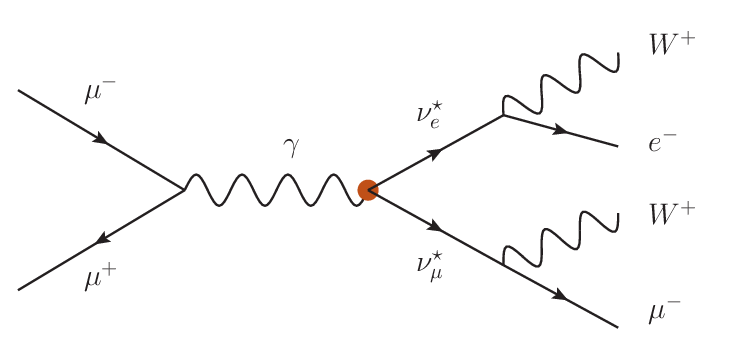}
 \caption{\em Leading (on the left) and sub-leading (on the right) Feynman diagrams for the $\Delta L=2$ process induced by an electron-muon neutrino dipole moment at a muon collider. The superscript $\star$ indicates that the exchanged neutrinos are off-shell.
 There is an equal contribution to the charge-conjugate process $\mu^+\mu^- \to W^- W^- e^+ \mu^+$.}\label{fig:NMM-muCol}
 \end{figure}

In order to probe the neutrino TDMs at a muon collider,
we identify as an efficient channel  
the $\Delta L=2$ process in Fig.~\ref{fig:NMM-muCol}, with two same-sign, different-flavour leptons, accompanied by two same-sign $W$'s, which is induced directly by the TDM operator in Eq.~\eqref{eq:NMM-A}. We focus again on the electron-muon TDM. In principle, a signature $\ell^\pm\tau^\pm$ can be used to probe electron-tau and muon-tau TDMs. However, we expect a lower sensitivities on these latter TDMs, due to a more difficult identification of the taus.  
Note that we are considering processes with the exchange of virtual, off-shell neutrinos.
The high collision energy of a muon collider, together with the derivative present in the effective photon-neutrino vertex of the TDM operator, allow for a 
significant cross section even for these processes. 
The more convenient final state to search for, because of the lower (reducible) background and the larger signal branching ratios, is the one with the two $W$'s decaying hadronically. This also permits a clear identification of the signal, with just two same-sign leptons in the final state.

Also in this case, we include the neutrino TDM interaction in Eq.~\eqref{eq:NMM-A} in {\tt MadGraph }5 \cite{Alwall:2014hca} by using {\tt Feynrules} \cite{Alloul:2013bka}, and calculate the cross section at a muon collider for several possible nominal collision energies,
 \[
 \sqrt{s} = 3, \, 10, \, 20, \, 30, \, 50 \, \text{TeV} \,.
 \]
 Signal and background events are simulated with {\tt MadGraph }5. Events are then passed to {\tt Pythia8} \cite{Bierlich:2022pfr} for showering.\footnote{Note that we also include QED shower, which gives non-negligible effects in the evaluation of the background.} We also apply a smearing to the jet 4-momenta, following the {\tt Delphes} \cite{deFavereau:2013fsa} default card, in order to minimally take into account detector effects.

\begin{figure}
\centering
 \includegraphics[scale=0.3]{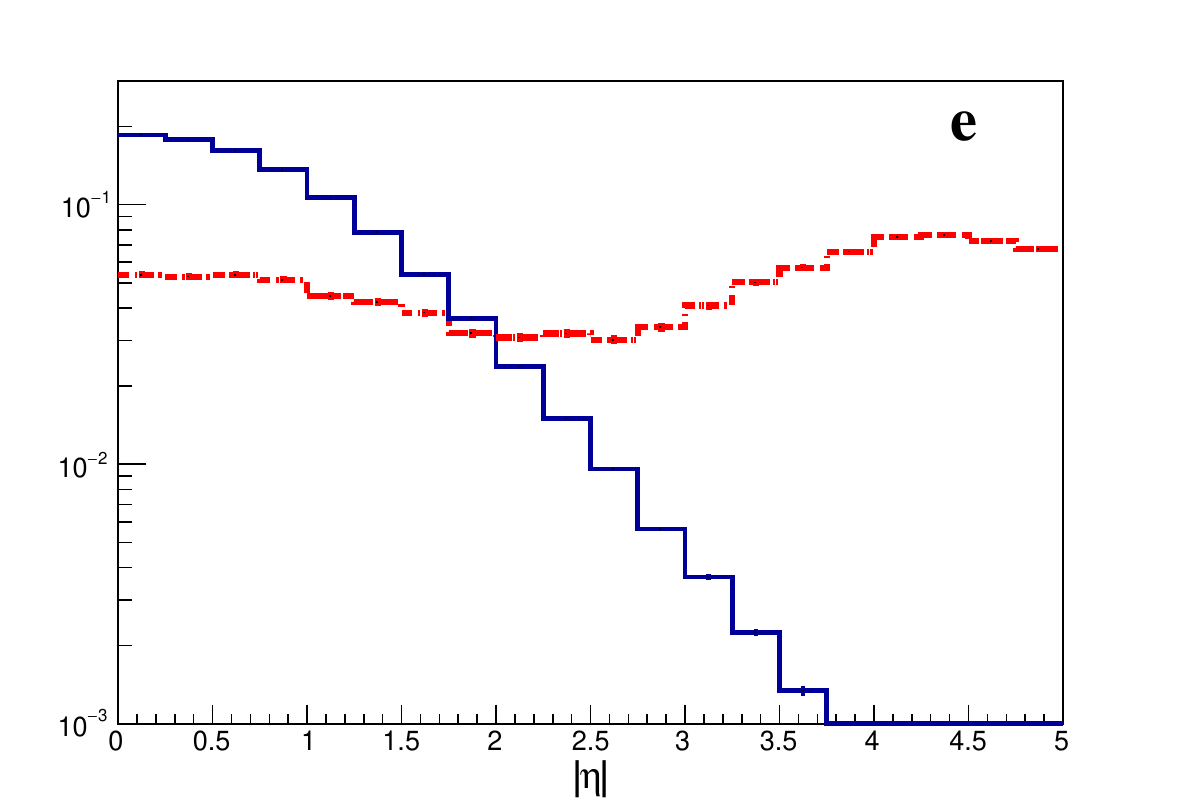} \includegraphics[scale=0.3]{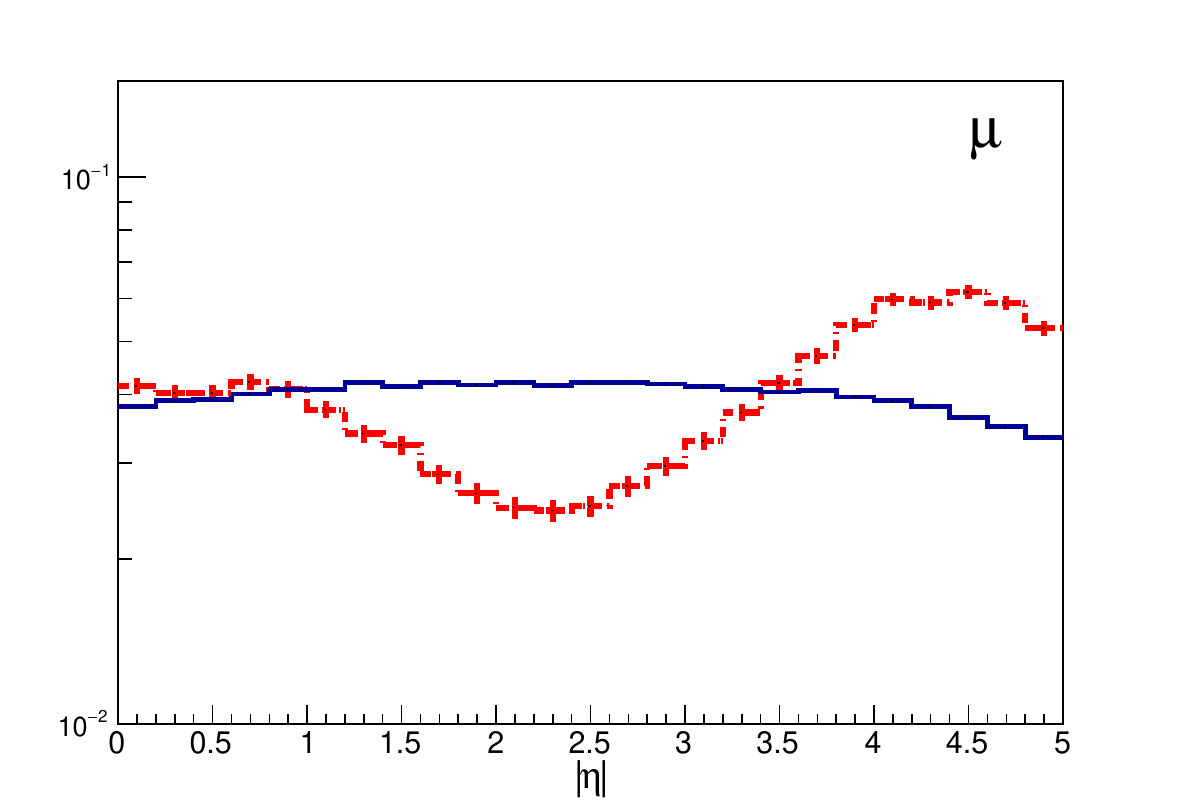}\\
 \includegraphics[scale=0.3]{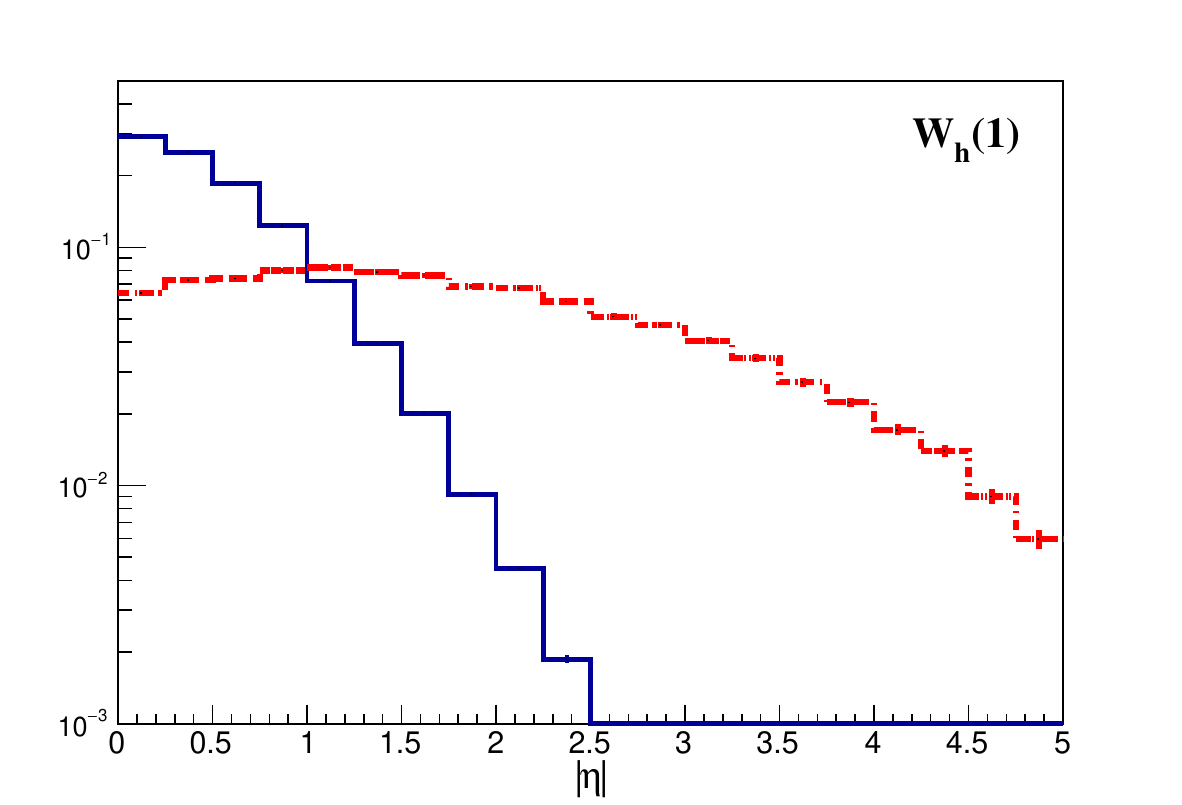} \includegraphics[scale=0.3]{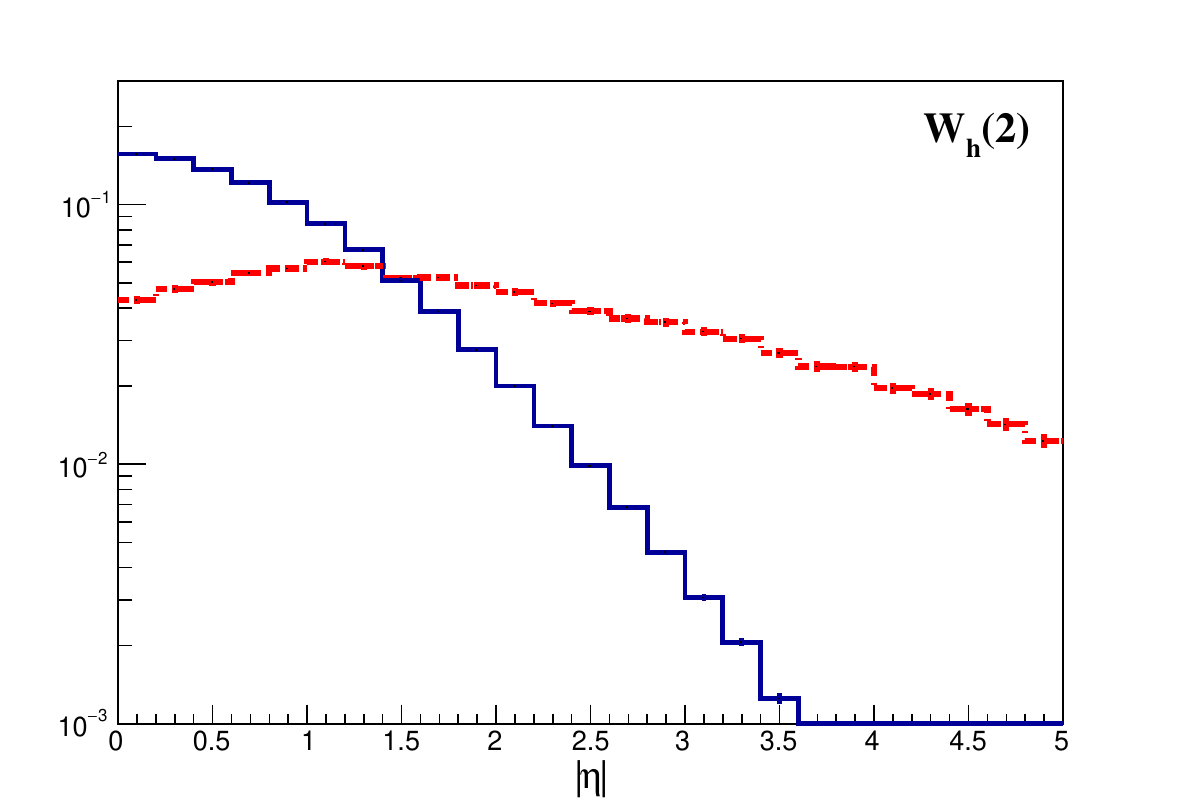}
 \caption{\em Rapidity distribution for the electron, the muon, the leading-$p_T$ $W_h$ and the second-leading-$p_T$ $W_h$ for the signal, in blue, and the total background, in dashed red. The distributions are normalized to a unit area and refer to a muon collider with $\sqrt{s}=20$ TeV. }\label{fig:eta}
 \end{figure}

We apply acceptance cuts of 20 GeV on the $p_T$ of the two leptons and the jets in the final state and we require the leptons and the jets to lie in the detector region, with a rapidity $|\eta|<5$. Shrinking the acceptance of the final muon to the central region, $|\eta|<2.5$, would indeed decrease by almost a factor of 2 the signal cross section.\footnote{The final muon is in fact emitted at a relatively large rapidity, as characteristic of a $t$-channel process of the type of the signal in Fig. \ref{fig:NMM-muCol} (left diagram). This is evident also from the muon rapidity distribution shown in Fig. \ref{fig:eta}. } We require the jets and the leptons to be separated by an angular distance $\Delta R_{j\ell}>$0.4, with $\Delta R=\sqrt{\Delta \eta^2+\Delta \phi^2}$. We also consider, conservatively, an identification efficiency of 85\% for both muons and electrons in the final state \cite{MuonCollider:2022ded}.
Each of the final state $W$ is produced with a large boost
 so that its hadronic decay products can be efficiently collected into a single fat-jet. We reconstruct the $W$-originated fat-jets with {\tt FastJet} \cite{Cacciari:2011ma} by using an anti-kt algorithm with cone size $R=0.5$. We find $W_h$ reconstruction efficiency of 93\% already at $\sqrt{s}=3$ TeV and approaching the 100\% efficiency at higher collision energies.

\begin{figure}
\centering
 \includegraphics[scale=0.7]{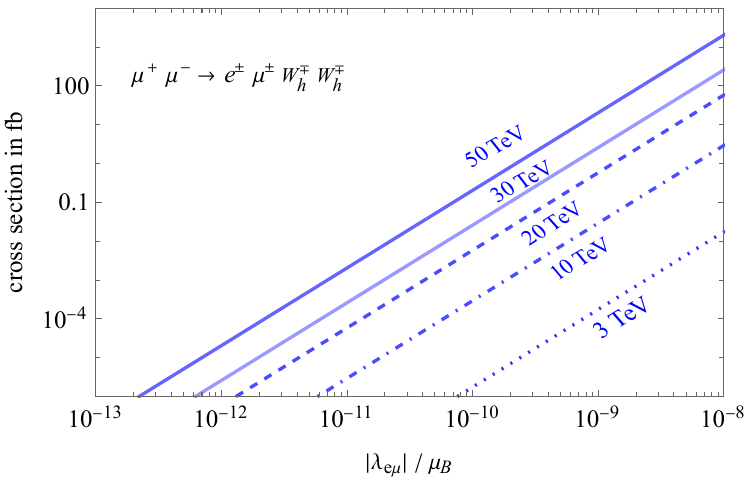}
 \caption{\em  Cross section values at a muon collider with different collision energies for the process $\mu^+ \mu^- \to e^\pm \mu^\pm W^\mp_h W^\mp_h$, as a function of the modulus of the $\nu_\mu-\nu_e$ TDM, in units of the Bohr magneton. }\label{fig:xsec-MuCol}
 \end{figure}

Cross section values we obtain, as function of the $\lambda_{e\mu}$ absolute value (expressed in units of $\mu_B$) are shown in Fig.~\ref{fig:xsec-MuCol}. Notice that a 3 TeV muon collider would already surpass the FCC-hh yield. The dominant contribution to the signal cross section is given by the $t$-channel process on the left of Fig.~\ref{fig:NMM-muCol}, which makes approximately the 90\% of the total signal cross section.
This is mainly due to the general $\sqrt{s}/E_V$ enhancement, with $E_V\ll \sqrt{s}$ the energy of the exchanged EW boson, of the cross section for processes with $t$-channel boson exchange, compared to $s$-channel ones (see for example \cite{Accettura:2023ked, Costantini:2020stv} and references therein). 

The signal we are considering, with a single pair of same-sign and different-flavour leptons accompanied by two hadronically decayed $W$'s, $W_h$, is very distinctive, since there are no processes in the SM which can reproduce the same final state. Therefore, the background would consist of reducible components and will be at a very low level at a muon collider. It would mainly arise  from $W_h W_h \ell^\pm \ell^\mp$ events, dominated by the muonic component $W_h W_h \mu^\pm \mu^\mp$,
where both the flavour and the charge of one of the final leptons is misidentified. Misidentification rates at a muon collider have been estimated to be less than 0.5\% \cite{Belfkir:2023lot, Yu:2017mpx} for the lepton flavour and about 0.1\% for the lepton charge \cite{ATLAS:2019jvq, Liu:2022kid}.\footnote{ The lepton charge misidentification rate is estimated on the basis of hadron collider detector performances.} An additional background source is generated by four-$W$ production processes, $W^\pm_h W^\pm_h W^\mp W^\mp$, where two $W$'s decay hadronically and two like-sign $W$'s decay leptonically, leading to a final state with two $W_h$ plus a $e^\pm \mu^\pm$ lepton pair, accompanied by large missing energy ($E_{miss}$). 
This background component can be reduced by applying a requirement on the maximally allowed missing energy. In particular, in our analysis we will exclude events with $E_{miss}>5\% \cdot \sqrt{s}$.\footnote{This cut is chosen conservatively based on the achievable detector performances of a muon collider \cite{MuonCollider:2022glg}.} We verify that a potential source of background coming from four-boson production processes, $W_h W_h VV$, with $V=\gamma^\star,Z$, where two of the final four leptons are missed, because either they fall outside the detector region or do not satisfy the minimum $p_T$ requirement, is negligible. Background components coming from tau leptonic decays and from $t\bar{t}(V)$ events are also negligible. Therefore, we do not include these potential sources in the $W_hW_h\ell\ell$ sample.
We evaluate a total background at the level of $\mathcal{O}(1)$ ab. 
 We report in Table \ref{tab:B} the background cross section values for the different collision energies. 

\begin{table}[]
\centering
{\footnotesize
\begin{tabular}{|c|c|c|c|c|c|}
\hline 
 & & & & &  \\
 & & & & &  \\[-0.6cm]
\textsf{$\sqrt{s}$}  & 3 TeV & 10 TeV & 20 TeV & 30 TeV & 50 TeV  \\  [0.2cm]
\hline
 & & & & & \\
 & & & & &  \\[-0.6cm]
  $W_h W_h \ell \ell$      &    0.57 (0.22) ab  & 1.0 (0.091) ab & 0.90 (0.044) ab & 0.73 (0.017) ab & 0.49 (0.0044) ab \\[0.17cm]
  $W_h W_h WW$        & 0.85 (0.57) ab & 0.92 (0.38) ab & 0.65 (0.22) ab & 0.49 (0.15) ab & 0.33 (0.092) ab \\[0.17cm]
  total                  & &  &&&\\
  background      & 1.4 (0.79) ab  &  1.9 (0.47) ab  & 1.6 (0.26) ab & 1.2 (0.17) ab & 0.82 (0.096) ab\\[0.15cm]
\hline
\end{tabular}
}
\caption{\label{tab:B}
\em  Cross section values for the background, after acceptance requirements and (values in parenthesis) the additional selection cuts in Eq.~\eqref{eq:cuts}.}
\end{table}

\begin{figure}
\centering
 \includegraphics[scale=0.29]{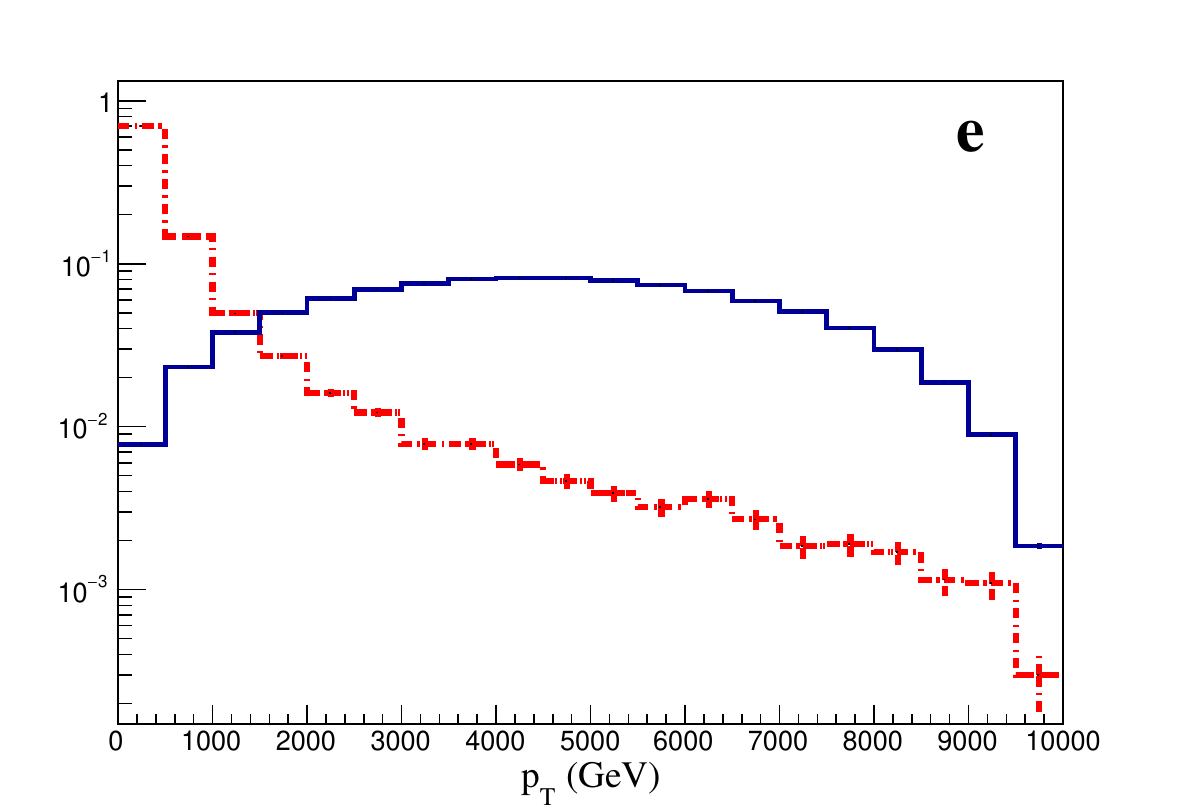} \includegraphics[scale=0.3]{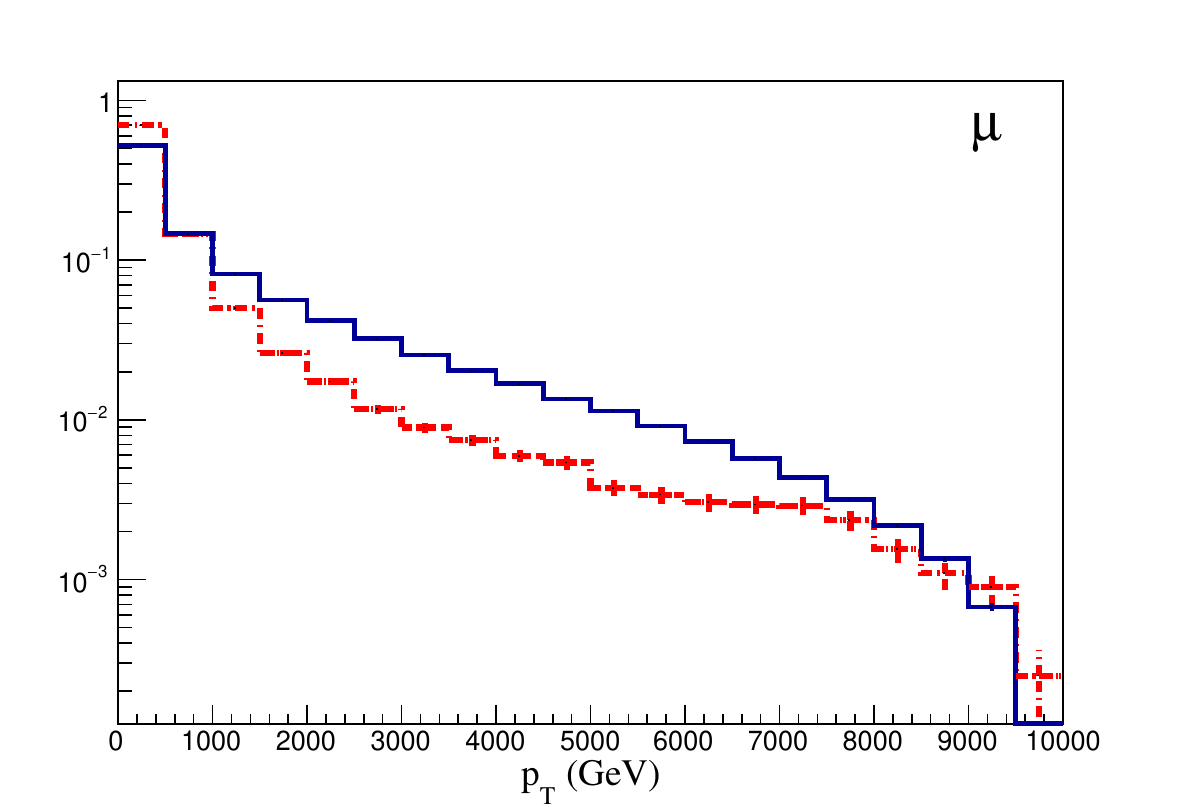}\\
 \includegraphics[scale=0.3]{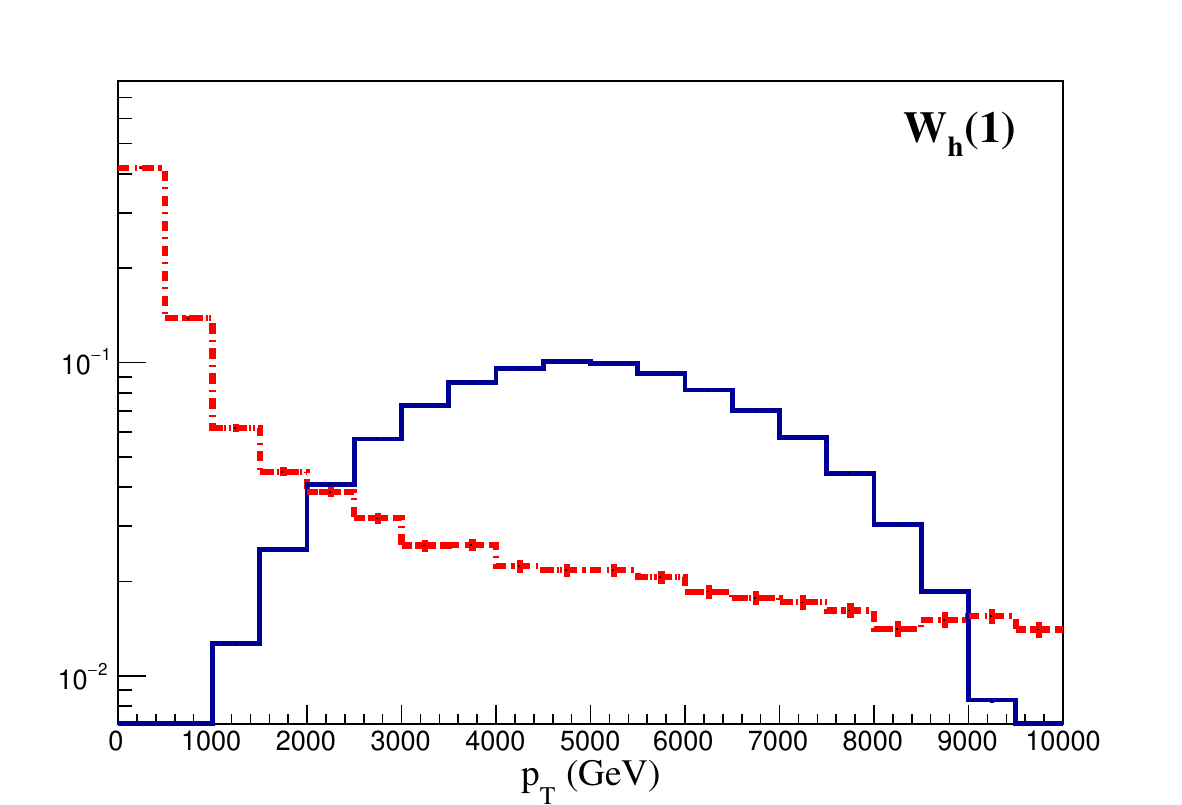} \includegraphics[scale=0.3]{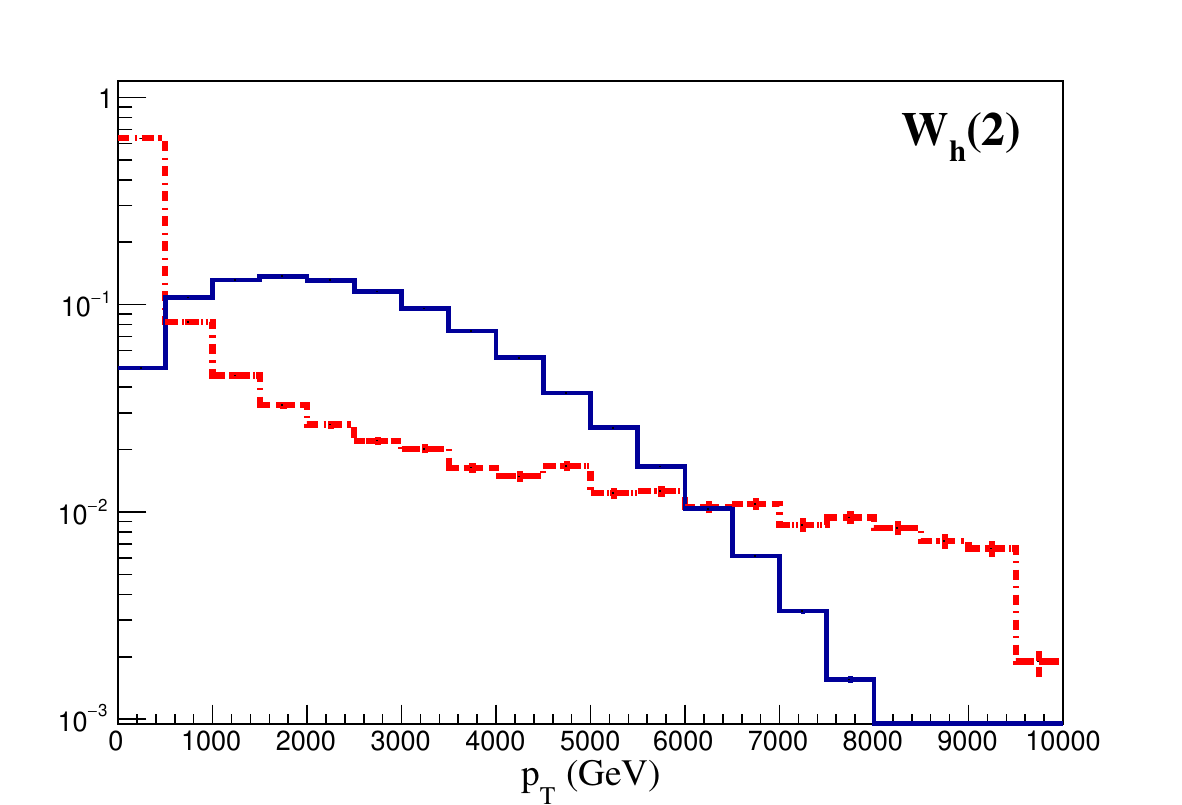}
 \caption{\em Transverse momentum distribution for the electron, the muon, the leading-$p_T$ $W_h$ and the second-leading-$p_T$ $W_h$ for the signal, in blue, and the total background, in dashed red. The distributions are normalized to a unit area, and refer to a muon collider with $\sqrt{s}=20$ TeV. }\label{fig:pt}
 \end{figure}

The signal we are focusing on is also characterized by a peculiar kinematics, which can be exploited to further reduce the background and identify the signal.
 We show in Fig.~\ref{fig:pt} the transverse momentum distributions and in Fig.~\ref{fig:eta} the rapidity distributions for the final state particles: the electron, the muon, and the two $W_h$'s, distinguished based on their $p_T$: $W_h(1)$ is the boson with the highest $p_T$ in the event, $W_h(2)$ the one with the lowest.\footnote{Conservatively, we do not apply restrictions on the background based on the reconstruction of the two $W_h$'s. For the background, in this analysis the two $W_h$'s are identified based on Monte Carlo truth. In principle, the requirement of only two fat-jets in the final state, reconstructed for example 
 by an anti-kt algorithm with cone-size $R=0.5$, would lead to a very efficient reconstruction of the $W_h$'s in the signal. 
 For the background, however, the two $W_h$'s have a boost which is, although still high, generally lower than in the signal. Therefore, 
 the same reconstruction procedure would be less efficient, possibly leading to a significant background reduction. Additionally, the resolution of the $V$ hadronic decays, which would be efficient for collision energies up to $\sim$10 TeV, would help in attenuating the background.
   A precise evaluation of these effects would need a refined analysis of the inclusive $\ell\ell+jets$ background, including potential misidentification of $W_h$'s from $\ell\ell VV$ or $\ell\ell VVV$ events, which goes beyond the scope of the present study.  }  
The signal is characterized by a final state electron and $W_h$'s emitted at high-$p_T$, much higher than those from the background, and mostly in the central region, while background final particles tend to be emitted with a relatively higher rapidity, in particular the component $W_hW_h\ell\ell $, which is dominated by EW boson scattering processes characterized by forward (high rapidity) emission of low-energy final state muons (one of which is misidentified).  
Similarly, for the signal, the final muon, as typical of the $t$-channel topology of the dominant signal process, is characterized by a relatively lower $p_T$ and a higher rapidity.      

In order to refine our analysis, we apply the following set of minimal cuts on the $p_T$ of the electron and of the leading-$p_T$ $W_h$:
\begin{equation}\label{eq:cuts}
p_T e > 2.5\%\cdot\sqrt{s}\,, \qquad p_T W_h(1) > 5\%\cdot\sqrt{s}\,.
\end{equation}
The signal efficiency to this selection is of 98\% at $\sqrt{s}=3$ TeV and approaches 100\% for higher collision energies, while the background, in particular the $W_hW_h \ell\ell$ component from lepton misidentification, is significantly reduced. Background cross section values after this selection are reported (values in parenthesis) in Table \ref{tab:B}. The selection we apply is rather simple and conservative, we therefore expect a significant improvement in the analysis strategies for a future experiment, which could for example exploit all the different kinematic distributions and advanced boosted-decision-tree (BDT) techniques.

Taking into account the maximal achievable integrated luminosity at a future muon collider \cite{Accettura:2023ked, Han:2020uak}, 
\begin{equation}
\mathcal{L} = 10 \left( \frac{\sqrt{s}}{10 \, \text{TeV}} \right)^2 \text{ab}^{-1}  \, ,    
\end{equation}
which is expected to be obtained in a 5 years data-taking time  \cite{Accettura:2023ked}, we can estimate the 2$\sigma$ sensitivities to $|\lambda_{e\mu}|$, as given in Table \ref{sensMU}. 
In the square brackets we also report the sensitivity values that could be obtained in a more optimistic, but still realistic, scenario, where the background is reduced to a negligible level and slightly better lepton identification efficiencies are considered, 90\% instead of 85\%. 
In our analysis we do not consider the impact of systematics, which will clearly be precisely assessable only once the experiment is actually operational. We emphasise however that the research channel we propose, enjoying a sizable signal-to-background ratio, is generally little subject to systematic uncertainties. 
We can see that the current best laboratory bound, $|\lambda_{e\mu}|\lesssim 6 \cdot 10^{-12}\mu_B$, can be matched by a muon collider with $\sqrt{s} \simeq 30$ TeV. Sensitivities can improve by about 40\% in the more optimistic scenario.  
As a final remark, we discuss how much the sensitivities would change if the detector geometry were to be restricted to the central region only, due to potential problems associated to the beam-induced-background. Limiting our analysis to an acceptance region $|\eta|\lesssim 2.7$, we find that the signal would be reduced by about 40\%, and a similar attenuation would be observed for the background. This would result in a variation of less than 15\% on the sensitivity values displayed above.

For illustrative purposes, we can denote the Wilson coefficient of the dim-5 TDM operator as an inverse energy scale, $|\lambda_{e\mu}|\equiv 1/\Lambda_{e\mu}$. The sensitivities found and reported in Table 2 then correspond to the exploration of an interval 
$10^3$ TeV $\lesssim \Lambda_{e\mu}\lesssim 10^6$ TeV. However, due to various suppression factors, the actual scale of new physics must be lower. In particular, as explained in section \ref{sec:theo}, the dim-5 TDM operator is the low-energy realisation of the dim-7 operators $O_B$ and $O_W$,
and the correspondent, more refined, estimate of $\lambda_{e\mu}$  is shown in Eq.~\eqref{NDA}.

\begin{table}[]
\centering
{\small
\begin{tabular}{|c|ccccc|}
\hline 
 & & & & &  \\
 & & & & &  \\[-0.6cm]
\textsf{$\sqrt{s}$} & 3 TeV & 10 TeV & 20 TeV & 30 TeV & 50 TeV \\
  \hline \\[-2mm]   {\Large $\frac{|\lambda_{e\mu}|}{ \mu_B} $}  & 5.4\;[4.8] $\cdot$ 10$^{-9}$ & 1.5\;[1.1]  $\cdot$ 10$^{-10}$ &  1.9\;[1.2]  $\cdot$ 10$^{-11}$ & 6.6\;[3.9]  $\cdot$ 10$^{-12}$ & 1.5\;[0.8]  $\cdot$ 10$^{-12}$ \\[0.15cm]
\hline
\end{tabular}
}
\caption{\label{sensMU}
\em The $2\sigma$ sensitivity limit on the $\nu_e-\nu_\mu$ TDM at a muon collider, according to our conservative analysis and, values in the brackets, for a more optimistic scenario, where the background is reduced to a negligible level and slightly better lepton identification efficiencies are considered. 
}
\end{table}

\subsection{Probing the effective operators $O_B$ and $O_W$  
}

In the previous section we considered a $\nu_e-\nu_\mu$ TDM in isolation. Here we will take into account the embedding of the TDM into the operators $O_B$ and $O_W$, defined in Eqs. \eqref{eq:OB} and \eqref{eq:OW}.

\begin{figure}[t]
\centering
 \includegraphics[scale=0.55]{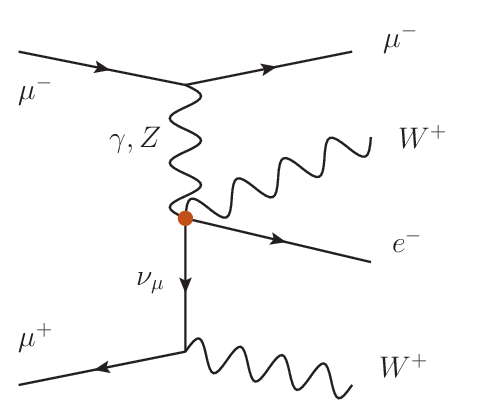} \includegraphics[scale=0.55]{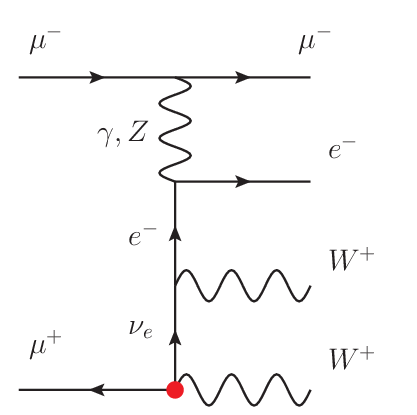}
 \includegraphics[scale=0.55]{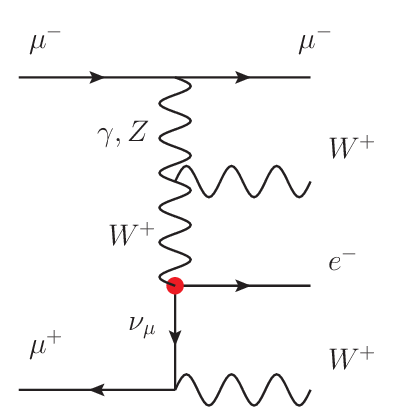}
 \caption{\em Illustrative Feynman diagrams for the $\Delta L=2$ process generated by two-bosons (first diagram on the left) and single-$W$ effective interactions (second and third diagrams) induced by the $O_W$ operator. The two types of contributions interfere destructively. }\label{fig:OW-diagram}
 \end{figure}

Indeed, once $SU(2)_w\times U(1)_Y$ invariance is imposed,   other effective interactions emerge from the $O_B$ and $O_W$ operators, as displayed in Eqs.~\eqref{eq:OB-interaction} and ~\eqref{eq:OW-interaction}, which lead to additional contributions to the process $\mu^{+} \mu^{-} \to e^{\pm}\mu^\pm W_h^\mp W_h^\mp$. For the $O_B$ operator, beside the contribution to the neutrino TDM effective vertex with a photon, there is an analogous vertex with the $Z$ boson: the Feynman diagrams are analogous to those in Fig.~\ref{fig:NMM-muCol} with the photon replaced by the $Z$. For the $O_W$ operator, beside the $Z$ and $\gamma$ interactions with two neutrinos, there are effective vertices with a $W$, a neutrino and a charged lepton (with or without an additional neutral gauge boson), as shown in the second line of Eq.~\eqref{eq:OW-interaction}. These lead to contributions to the LNV process we are analysing, with a few representative Feynman diagrams shown in Fig.~\ref{fig:OW-diagram}.\footnote{We obtain a total of 144 diagrams for the process $\mu^+\mu^-\to e^\pm \mu^\pm W_h ^\mp W_h^\mp$ induced by $(O_W)_{e\mu}$. Specifically, 64 $s$-channel diagrams (which are subdominant), 72 $t$-channel processes from effective single-$W$, $W\gamma$, and $WZ$ interactions (which include the diagrams in Fig.~\ref{fig:OW-diagram}), and 8 diagrams from single-photon and single-$Z$ interactions (analogous to the diagrams in Fig.~\ref{fig:NMM-muCol}).} 

Considered individually, the two types of contribution, those from a $W\gamma$ or $WZ$ vertex (first diagram in Fig.~\ref{fig:OW-diagram}) and those from a single-$W$ vertex (second and third diagrams in Fig.~\ref{fig:OW-diagram}), would give a strong increase to the cross section. However, we observe a destructive interference between the two types of contribution (note the former contains one less $t$-channel propagator, while the latter contains an extra momentum in the effective vertex). As a consequence, the cross section for $O_W$ is overall of the same size as that for the $O_B$ operator, with the dominant contribution actually coming from single-photon or $Z$ interactions. Indeed, before electroweak symmetry breaking one expects a comparable rate for $\Delta L = 2$ processes induced by $O_B$ and $O_W$, and this result has to persist once the details of electroweak symmetry breaking
are taken into account.

\begin{figure}
\centering
 \includegraphics[scale=0.7]{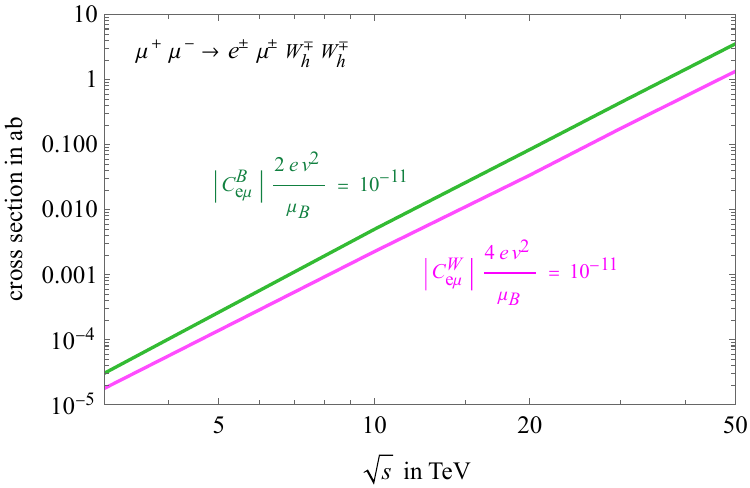}
\caption{\em Cross section values at a muon collider for the process $\mu^+ \mu^- \to e^\pm \mu^\pm W_h^\mp W_h^\mp$ as a function of the collision energy, for a reference value $10^{-11}$ of the quantities $|C^B_{e\mu}|\frac{2 e v^2}{ \mu_B} $ (with $C^W_{e\mu}=0$) and $|C^W_{e\mu}|\frac{4 e v^2}{ \mu_B} $ (with $C^B_{e\mu}=0$), corresponding to $|\lambda_{e\mu}|/\mu_B =10^{-11}$. Note that the cross sections depend  on the modulus of the Wilson coefficients quadratically. }
\label{fig:xsec-MuCol-Operators}
 \end{figure}

Cross section values at a muon collider for the process $\mu^+ \mu^- \to e^\pm \mu^\pm W_h^\mp W_h^\mp$, generated by the $(O_B)_{e\mu}$ and the $(O_W)_{e\mu}$ operators, are presented in Fig.~\ref{fig:xsec-MuCol-Operators}. Cross sections are shown as a function of the muon collider collision energy, and for a reference value of the Wilson coefficients $C^B_{e\mu}$ and $C^W_{e\mu}$, normalised to a dimensionless quantity which matches their contribution to $|\lambda_{e\mu}|/\mu_B$, according to Eq.~\eqref{match}. 

 Cross sections for the $O_B$ case are roughly a factor of two  higher than those from the purely TDM operator in Fig.~\ref{fig:xsec-MuCol}, while cross sections for the $O_W$ case are slightly lower. 
Kinematic distributions for the signal are very similar to those shown in Figs.~\ref{fig:pt} and \ref{fig:eta} for the case of neutrino TDM only, due to the analogous dominant topology (kinematic distributions for the background are obviously identical). Therefore, we can repeat the signal selection strategy applied for the purely TDM case. We obtain the  2$\sigma$ sensitivities presented in Table \ref{sens-CB-CW}. We conclude that a muon collider has a slightly higher  sensitivity to the $C^B_{e\mu}$ Wilson coefficient compared to $C^W_{e\mu}$.  
By comparing the sensitivities in Table \ref{sens-CB-CW} with the NDA estimate for the neutrino dipole in Eq.~(\ref{NDA}), one observes that the muon collider can test new physics with mass $m_*$ close to $\sqrt{s}$. Therefore, in the case of a positive signal, it might be interesting to go beyond the EFT approach to the neutrino dipole, as we further discuss in section \ref{sec:conclusions}.

\begin{table}[t]
\centering
{\footnotesize
\begin{tabular}{|c|ccccc|}
\hline 
 & & & & &  \\
 & & & & &  \\[-0.6cm]
\textsf{$\sqrt{s} $} & 3 TeV & 10 TeV & 20 TeV & 30 TeV & 50 TeV \\
  \hline 
   & & & & &  \\
 & & & & &  \\[-0.6cm]
  {\large $|C^B_{e\mu}|\frac{2 e v^2}{ \mu_B} $}  & 4.0\;[3.6] $\cdot$ 10$^{-9}$ & 1.2\;[0.9]  $\cdot$ 10$^{-10}$ &  1.6\;[1.0]  $\cdot$ 10$^{-11}$ & 5.0\;[3.0]  $\cdot$ 10$^{-12}$ & 1.2\;[0.6]  $\cdot$ 10$^{-12}$ \\
  \hline   & & & & &  \\
 & & & & &  \\[-0.6cm] {\large $|C^W_{e\mu}|\frac{4 e v^2}{ \mu_B} $}  & 5.4\;[4.8] $\cdot$ 10$^{-9}$ & 1.7\;[1.2]  $\cdot$ 10$^{-10}$ &  2.4\;[1.5]  $\cdot$ 10$^{-11}$ & 7.8\;[4.6]  $\cdot$ 10$^{-12}$ & 1.8\;[1.0]  $\cdot$ 10$^{-12}$ \\[0.15cm]
\hline
\end{tabular}}\caption{\label{sens-CB-CW}
\em The $2\sigma$ sensitivity limit on the quantities $|C^B_{e\mu}|\frac{2 e v^2}{ \mu_B} $ and $|C^W_{e\mu}|\frac{4 e v^2}{ \mu_B} $, which match $|\lambda_{e\mu}|/\mu_B$ when, respectively, $C^W_{e\mu}=0$ and $C^B_{e\mu}=0$,
as a function of the muon collider energy. The first number corresponds to our conservative analysis, while in brackets we report the sensitivity for a more optimistic scenario, where the background is reduced to a negligible level, and slightly better lepton identification efficiencies are assumed. 
}
\end{table}

\section{Testing neutrino masses at a muon collider }\label{sec:masses}

The LNV signatures here considered to probe neutrino TDMs at colliders are also sensitive to Majorana neutrino masses,

\begin{equation}\label{MajM}
{\cal L}\supset -\dfrac 12 \overline{(\nu_{L\alpha})^c} \, m_{\alpha\beta} \, \nu_{L\beta}+\text{h.c.}\,,
\end{equation}
with $m$ a complex symmetric $3\times 3$ matrix in flavour space.

Such Majorana mass term can be induced by the only dimension-5 invariant operator of the SM EFT, the so-called Weinberg operator \cite{Weinberg:1979sa}, 
\begin{equation}
\mathcal{L}_{5}=
C^5_{\alpha\beta}
\left(\overline{\ell_{L\alpha}^{\,c}}\epsilon H\right)  \left(H^T\epsilon \ell_{L\beta}\right)
+ {\text h.c.}\,,\qquad
m_{\alpha\beta}\equiv C^5_{\alpha\beta}v^2\,.
\end{equation}
As the neutrino TDMs, Majorana masses violate lepton number by two units. The only matrix element subject to a tight, direct experimental constraint is $m_{ee}$, with an upper bound from neutrinoless $2\beta$ decay searches of the order of 0.1 eV \cite{GERDA:2020xhi, KamLAND-Zen:2016pfg}. In the minimal scenario with three active Majorana
neutrinos only, oscillations experiments imply that all matrix elements $m_{\alpha\beta}$ should respect a similar upper bound. Nonetheless, it is interesting to conceive measurements which could be directly sensitive to $m_{\alpha\beta}$ for $\alpha\beta\ne ee$ to avoid theoretical prejudice and test non-minimal scenarios.\footnote{
For example, in the presence of light sterile neutrinos, one can define effective mass parameters $m_{\alpha\beta}$ which are enhanced by the sterile neutrino contribution, see e.g.~Ref.~\cite{Abada:2017jjx}.}

The same LNV signature $pp \to \ell^\pm_{\alpha} \ell^{\pm}_\beta jj$, considered in section~\ref{sec:hadron-colliders} to probe TDMs at hadron colliders, has been recently exploited as a channel to probe Majorana neutrino masses
\cite{Fuks:2020zbm}.
The analysis in \cite{Fuks:2020zbm} indicate projected sensitivities on $|m_{\mu\mu}|$ of $\sim$ 5.4 GeV at the HL-LHC with 3 ab$^{-1}$, and of $\sim$ 1.2 GeV at the FCC-hh with 30 ab$^{-1}$. The recent ATLAS searches \cite{ATLAS:2023tkz, ATLAS:2024rzi}, performed at the 13 TeV LHC with 140 fb$^{-1}$ on the basis of the study in \cite{Fuks:2020zbm}, give the following 95\% C.L. bounds on Majorana masses: $|m_{e\mu}|<13$ GeV, $|m_{ee}|<24$ GeV \cite{ ATLAS:2024rzi} and $|m_{\mu\mu}|< 16.7$ GeV \cite{ATLAS:2023tkz}.
Based on these experimental results and on the projected sensitivities of \cite{Fuks:2020zbm}, we can roughly estimate the following exclusion sensitivities:
\begin{align}
\begin{split}\label{eq:Weinberg-hadron}
\text{HL-LHC \{3 ab$^{-1}$\}:} \qquad &|m_{\mu\mu}|\sim 5.4\, \text{GeV} \quad |m_{ee}|\sim 7.8\, \text{GeV} \quad |m_{e\mu}|\sim 4.2\, \text{GeV} \,,\\
\text{FCC-hh \{30 ab$^{-1}$\}:} \qquad &|m_{\mu\mu}|\sim 1.2\, \text{GeV} \quad |m_{ee}|\sim 1.7\, \text{GeV} \quad |m_{e\mu}|\sim 0.93\, \text{GeV} \,.
\end{split}
\end{align}

A recent study in \cite{Li:2023lkl} also presents projected sensitivities on Majorana masses of a future {\it same-sign} muon collider, analysing vector boson scattering processes. The study reports sensitivities similar to those of the FCC-hh,  assuming a 30 TeV collision energy.

In this section we will highlight  a strategy to probe Majorana masses at a future $\mu^+\mu^-$ muon collider, and we will present the corresponding projected sensitivities. 
As the initial state consists of muons, the processes induced by a single neutrino-mass insertion are those involving either $m_{\mu e}$,
$m_{\mu\mu}$,  or $m_{\mu\tau}$. For simplicity, as usual, we will focus on final states without $\tau$ leptons.

\begin{figure}[t]
\centering
 \includegraphics[scale=0.5]{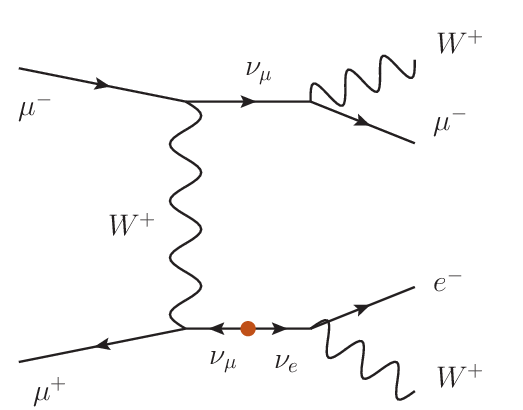}
 \includegraphics[scale=0.5]{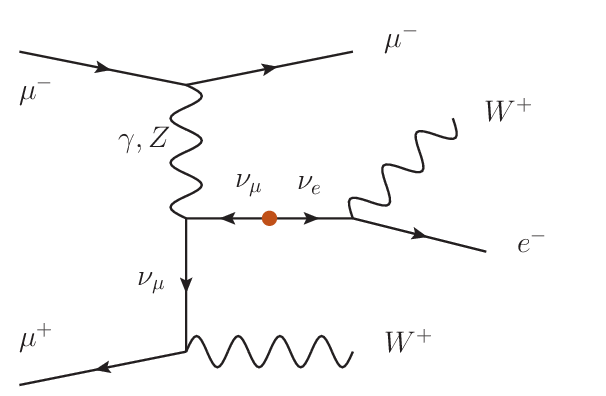} 
 \includegraphics[scale=0.5]{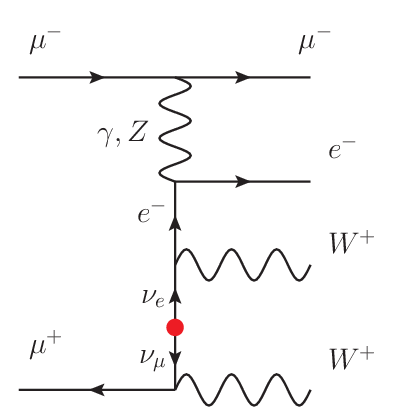}
 \caption{\em Representative leading Feynman diagrams for the $\Delta L=2$ process induced by a Majorana mass $m_{e\mu}$. 
 }\label{fig:wo-muCol}
 \end{figure}

The LNV processes $\mu^+ \mu^- \to \ell^\pm_\alpha \ell^{\pm}_\beta W^\mp_h W^\mp_h$, which we used to probe neutrino TDMs in the case $\ell_\alpha\ell_\beta\equiv e\mu$, can indeed test Majorana masses as well. Some representative leading diagrams for the Majorana-mass contribution to the process $\mu^+\mu^- \to e^\pm\mu^\pm W^\mp_hW^\mp_h$ are presented in Fig.~\ref{fig:wo-muCol}. 
We adopt the 
UFO {\tt MadGraph} implementation, {\tt SMWeinberg} \footnote{Available at https://{\tt Feynrules}.irmp.ucl.ac.be/wiki/SMWeinberg. We refer the reader to this web page and to Ref. \cite{Fuks:2020zbm} for details on the model implementation and usage. }, of the study in \cite{Fuks:2020zbm} for signal simulation. 

As in the previous analysis for the TDMs, we consider as acceptance criteria two same-sign leptons in the final state, with $p_T>$20 GeV and rapidity $|\eta|<$5, and two hadronically decayed $W$'s. Final state jets are also required to have a $p_T>$20 GeV and a rapidity $|\eta|<$5. The jets and the leptons must be separated by an angular distance $\Delta R_{j\ell}>$0.4. We consider an identification efficiency of 85\% for both muons and electrons in the final state and we exclude events with large missing energy, $E_{miss}>5$\%$\cdot\sqrt{s}$.

\begin{figure}[t]
\centering
 \includegraphics[scale=0.7]{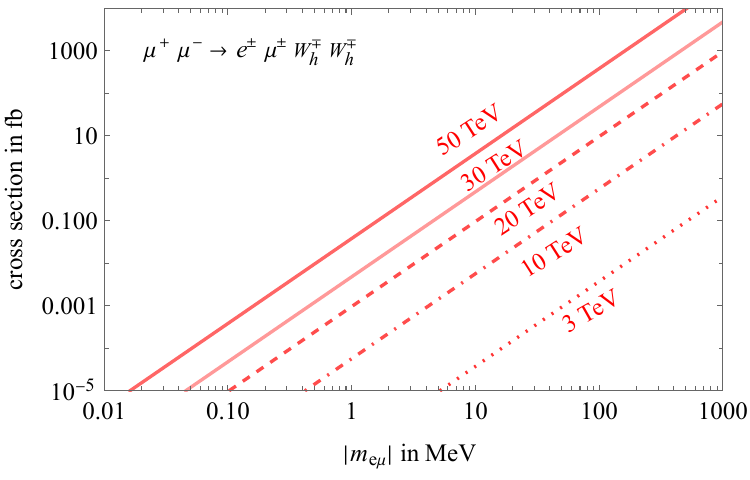}
\caption{\em Cross section values at a muon collider, with different collision energies, for the process $\mu^+\mu^- \to e^\pm\mu^\pm W^\mp_hW^\mp_h$, as a function of  $|m_{e\mu}|$.
 The cross section values are two times smaller for the process $\mu^+\mu^- \to \mu^\pm\mu^\pm W^\mp_hW^\mp_h$, as a function of $|m_{\mu\mu}|$. }\label{fig:WO-xsec-MuCol}
 \end{figure}

Fig.~\ref{fig:WO-xsec-MuCol} shows the cross section values for the Majorana-mass signal at a muon collider, as a function of the mass $|m_{e\mu}|$, 
for all other masses set to zero.  
Cross section values two times smaller (due to the symmetry factor for final state identical particles) are obtained for the process $\mu^+\mu^- \to \mu^\pm\mu^\pm W^\mp_hW^\mp_h$, when $m_{\mu\mu}$ 
is the only non-zero mass. 

\begin{figure}[tbp]
\centering
 \includegraphics[scale=0.3]{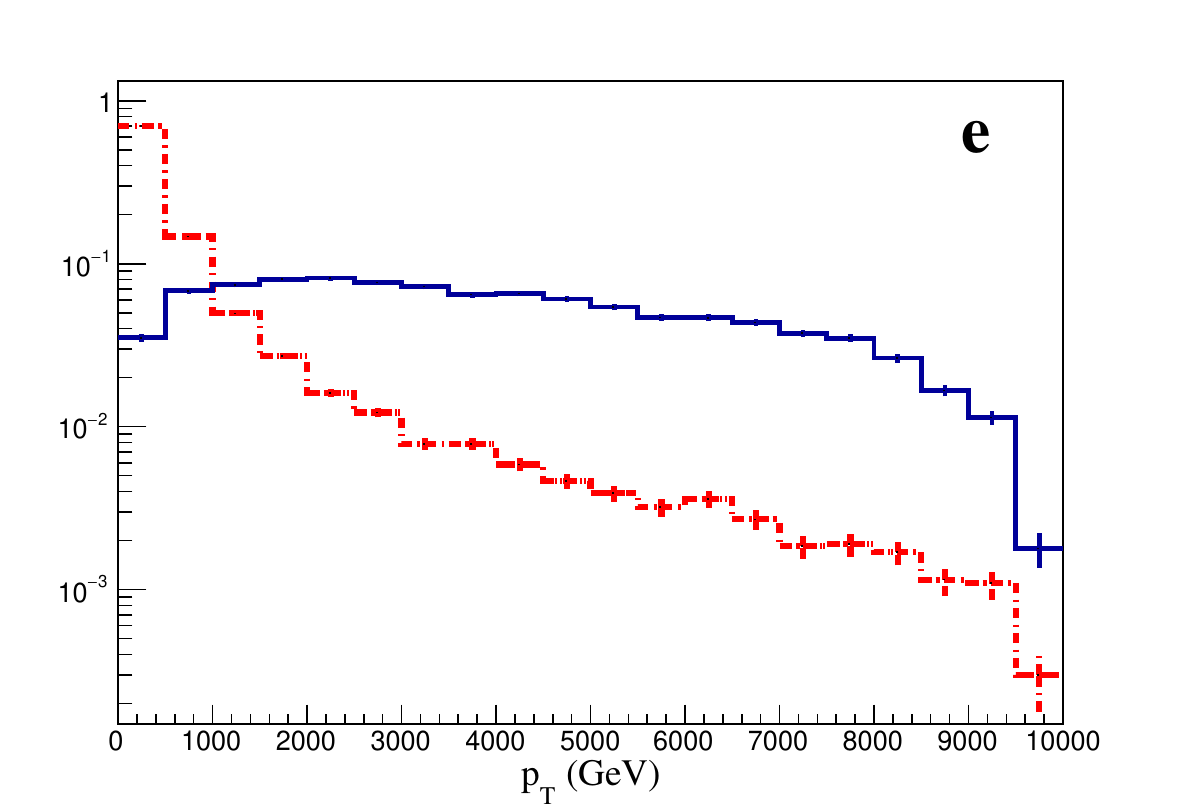} \includegraphics[scale=0.3]{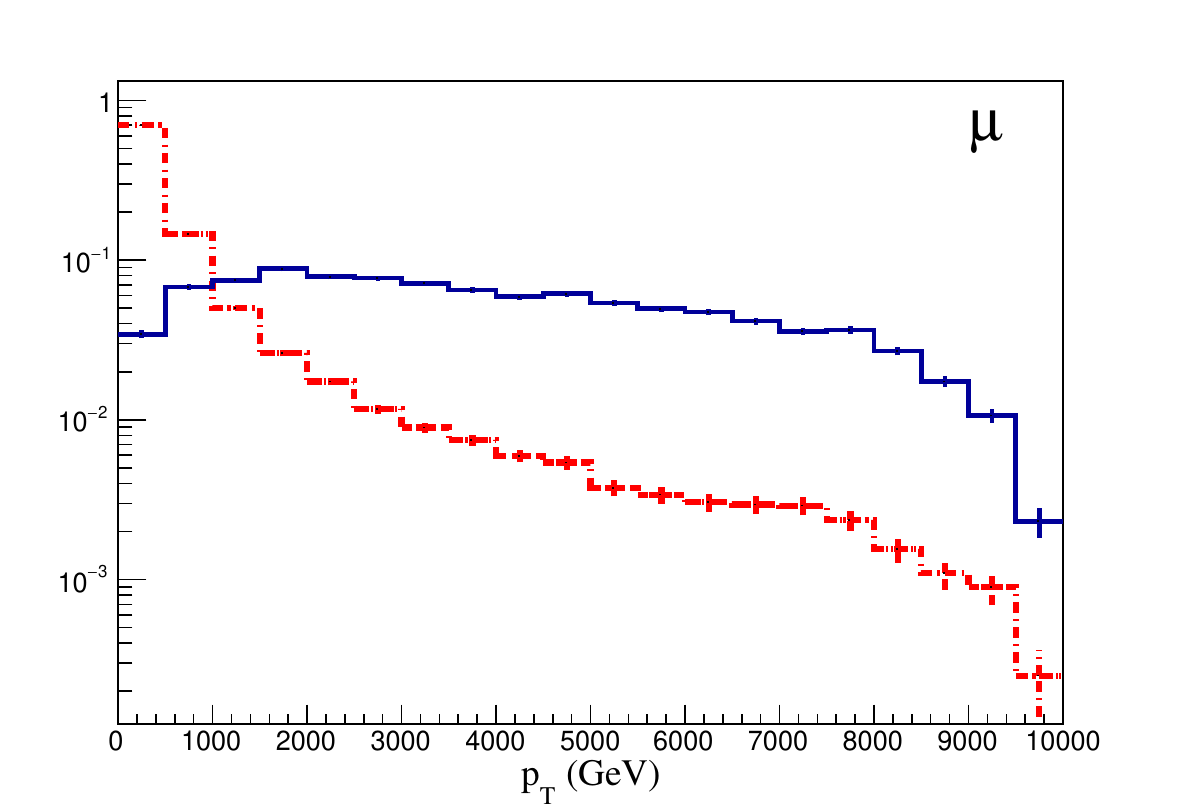}\\
 \includegraphics[scale=0.3]{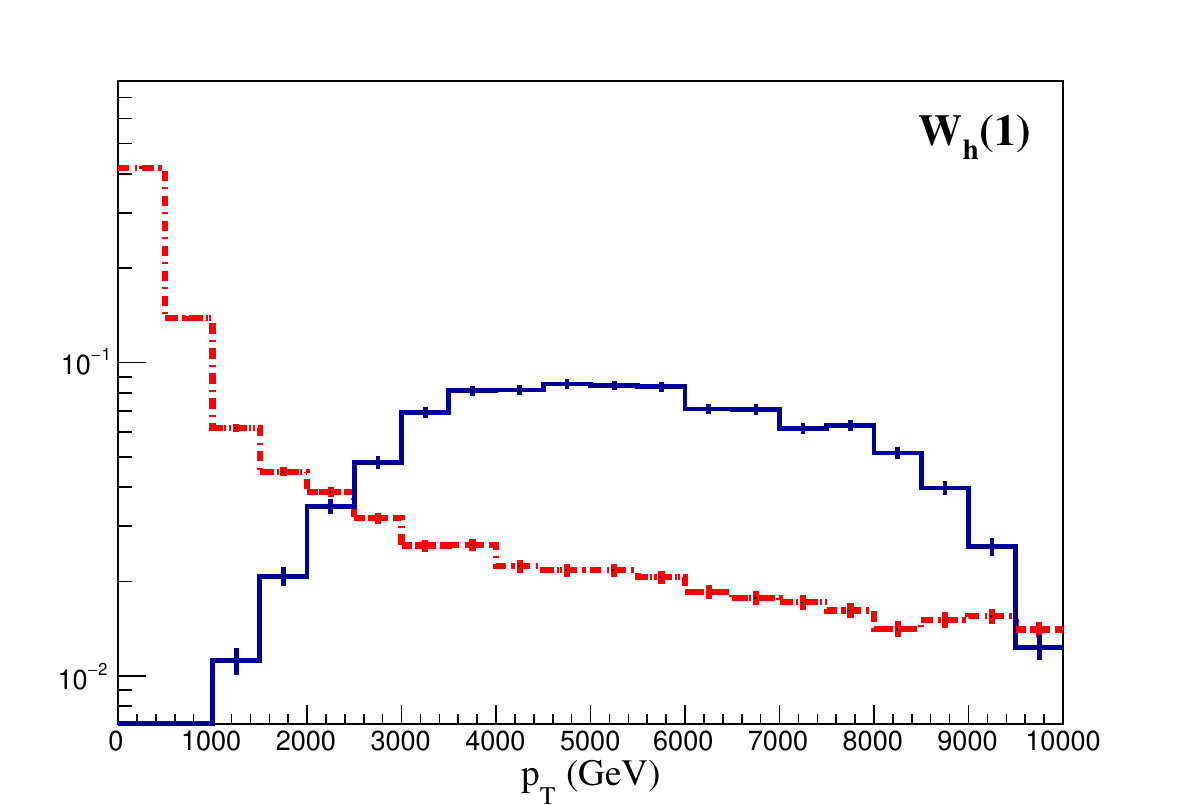} \includegraphics[scale=0.3]{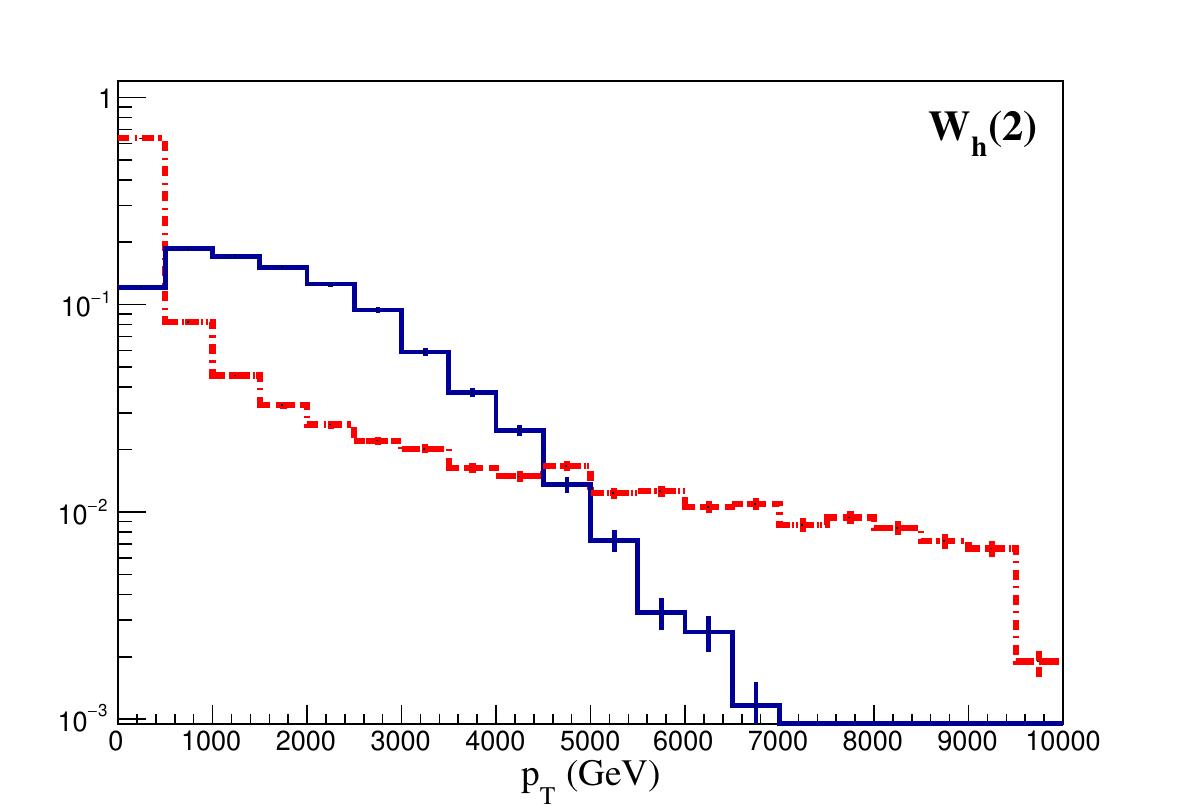}
 \caption{\em Transverse momentum distribution for the electron, the muon, the leading-$p_T$ $W_h$ and the second-leading-$p_T$ $W_h$ for the signal $\mu^+\mu^- \to e^\pm \mu^\pm W^\mp_h W^\mp_h$ induced by a Majorana mass $m_{e\mu}$, in blue, and the total background, in dashed red. The distributions are normalized to a unit area and refer to a muon collider with $\sqrt{s}=20$ TeV. }\label{fig:pt-ow}
 \end{figure}

 \begin{figure}[tbp]
\centering
 \includegraphics[scale=0.3]{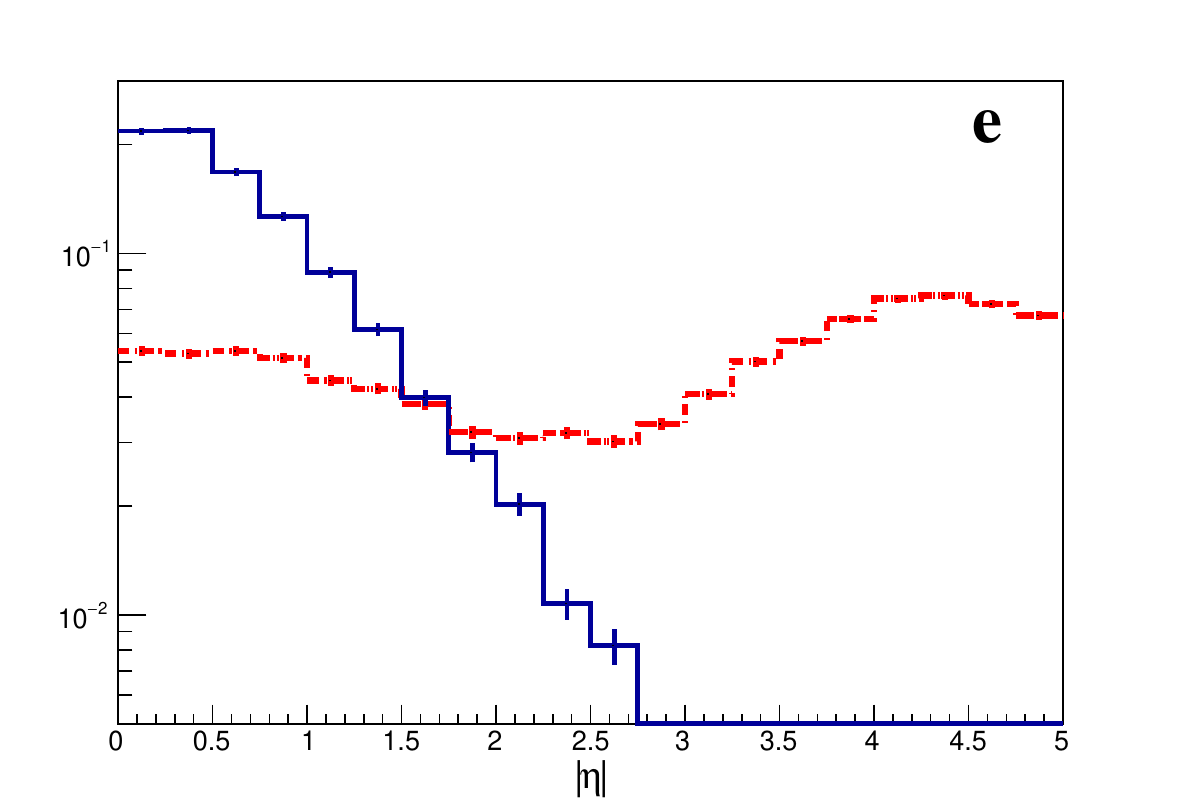} \includegraphics[scale=0.3]{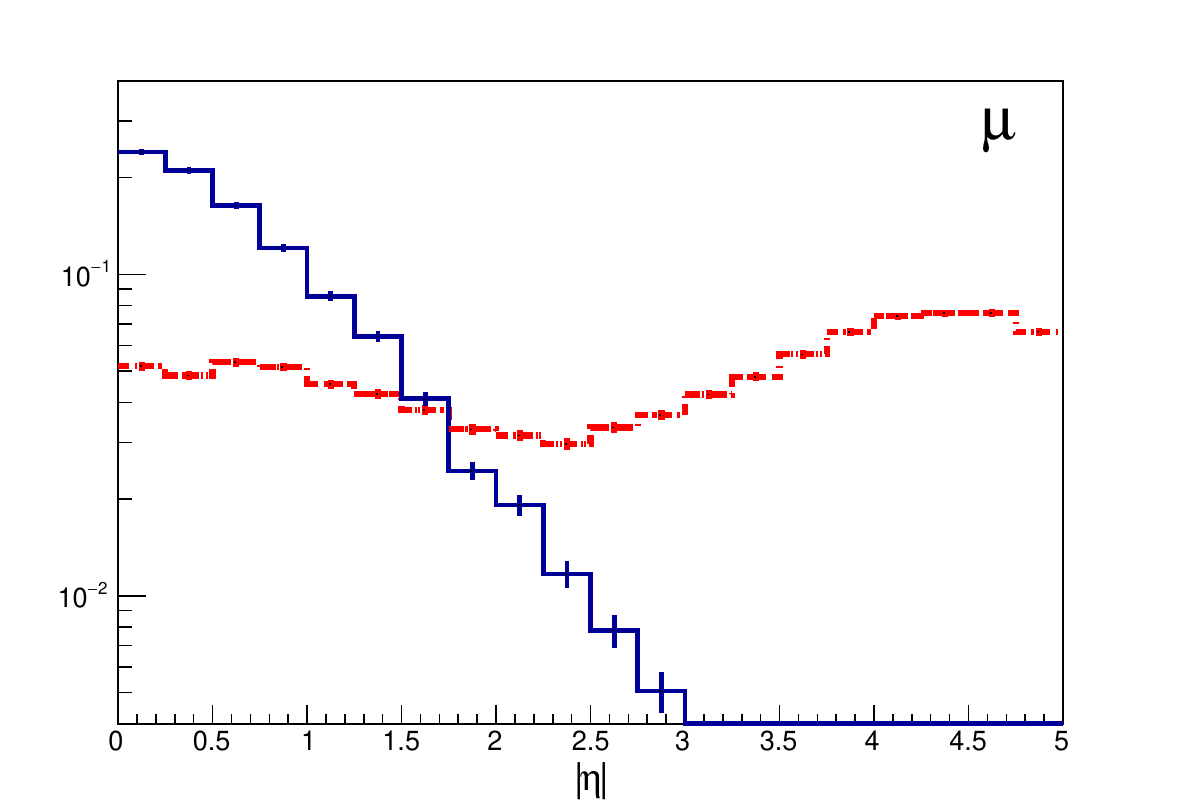}\\
 \includegraphics[scale=0.3]{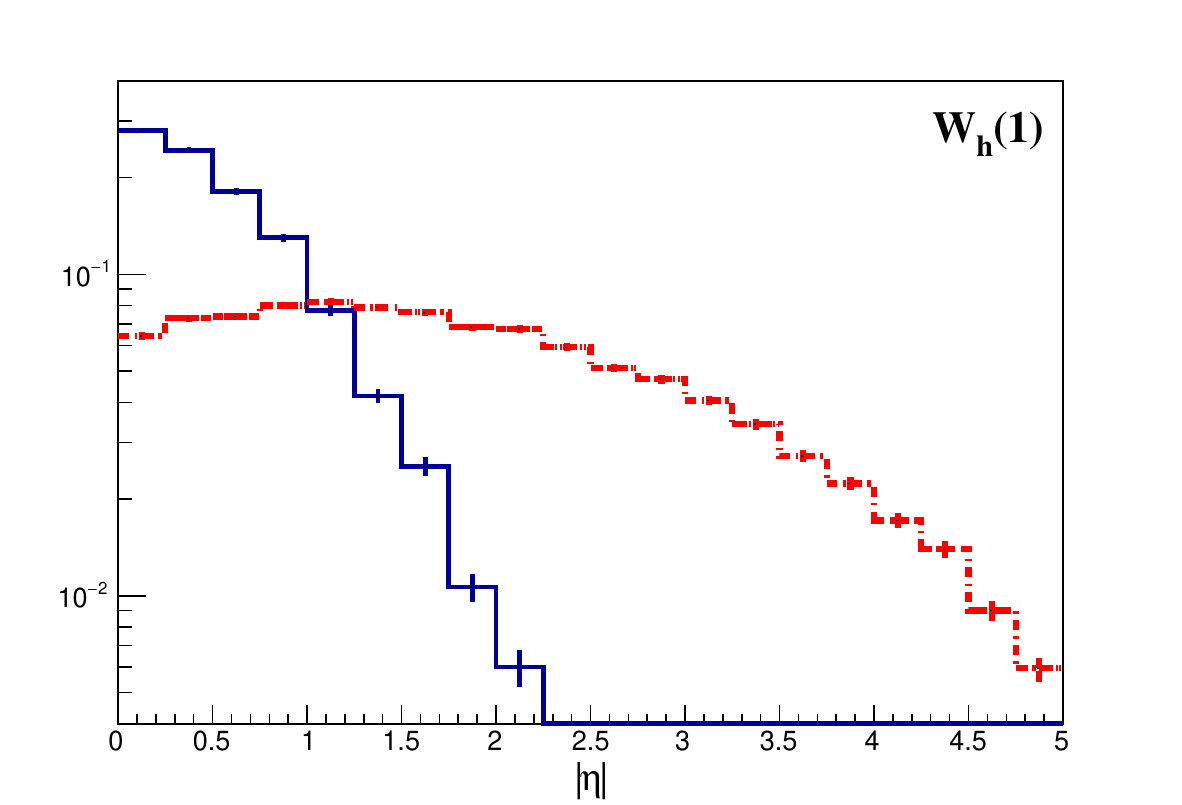} \includegraphics[scale=0.3]{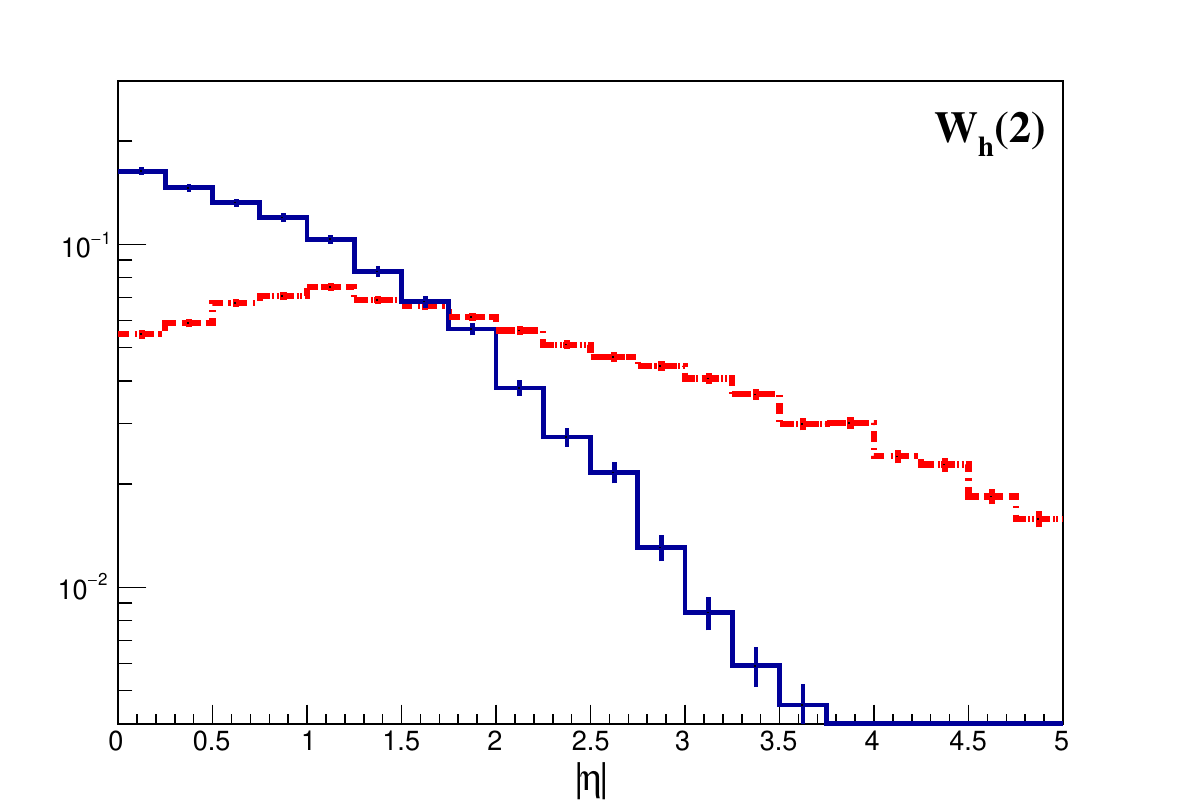}
 \caption{\em Rapidity distribution for the electron, the muon, the leading-$p_T$ $W_h$ and the second-leading-$p_T$ $W_h$ for the signal induced by a Majorana mass $m_{e\mu}$, in blue, and the total background, in dashed red. The distributions are normalized to a unit area and refer to a muon collider with $\sqrt{s}=20$ TeV.}\label{fig:eta-ow}
 \end{figure}

The kinematic distributions for the process $\mu^+\mu^- \to e^\pm \mu^\pm W^\mp_h W^\mp_h$, induced by the Majorana-mass operator, which we show in Figs.~\ref{fig:pt-ow} and \ref{fig:eta-ow}, are similar to those for the neutrino TDM case, shown in Figs.~\ref{fig:pt} and \ref{fig:eta}, with the only exception of the distributions of the final muon. Therefore, we can repeat the same selection strategy applied in the previous section for the neutrino TDM case to isolate a signal from the background. Notice that a subsequent analysis of the final muon $p_T$ and $\eta$ distributions could then provide a useful tool to characterize the signal and distinguish in particular the two cases: a TDM-induced signal versus a Majorana mass one. For the Majorana-mass case, in fact, the final muon tends to be emitted more centrally than in the TDM case, with $p_T$ and $\eta$ shapes analogous to those of the final electron.\footnote{For the Majorana mass signal, we can anticipate that an eventual restriction of the detector acceptance to the central rapidity region would lead to a slight increase in the sensitivity values. This is because signal events lie almost completely in the region $|\eta|\lesssim 2.7$, while the background extends more forward and would be thus reduced. Conservatively, we keep however our acceptance region at $|\eta|<5$.}
This is because various topologies contribute to the Majorana mass signal, giving a substantially symmetric kinematics for the electron and the muon, while a single $t$-channel diagram is dominant in the TDM case, leading to the typical distributions of the emitted muon in the final state.

\begin{table}[t]
\centering
{\footnotesize
\begin{tabular}{|c|c|c|c|c|c|}
\hline 
 & & & & &  \\
 & & & & &  \\[-0.6cm]
\textsf{$\sqrt{s}$}  & 3 TeV & 10 TeV & 20 TeV & 30 TeV & 50 TeV  \\  [0.2cm]
\hline
 & & & & & \\
 & & & & &  \\[-0.6cm]
  $W_h W_h \ell \ell$      &    57 (44) ab  & 100 (18) ab & 90 (9.0) ab & 73 (3.5) ab & 49 (0.97) ab \\[0.17cm]
  $W_h W_h WW$        & 0.9 (0.6) ab & 0.9 (0.4) ab & 0.7 (0.2) ab & 0.5 (0.1) ab & 0.3 (0.09) ab \\[0.17cm]
  Total                  & &  &&&\\
  background      & 58 (45) ab  &  101 (19) ab  & 91 (9.2) ab & 74 (3.6) ab & 50 (1.1) ab\\[0.15cm]
\hline
\end{tabular}
}
\caption{\label{tab:mumu-B}
\em  Background cross section values for the process $\mu^+\mu^-\to \mu^\pm\mu^\pm W^\mp_h W^\mp_h$, after acceptance requirements and, values in parentheses, after the additional cuts in Eq.~\eqref{eq:cuts-mumu}. }
\end{table}

\begin{table}[t]
\centering
{\small
\begin{tabular}{|c|ccccc|}
\hline 
 & & & & &  \\
 & & & & &  \\[-0.6cm]
\textsf{$\sqrt{s}$} & 3 TeV & 10 TeV & 20 TeV & 30 TeV & 50 TeV \\
  \hline  & & & & & \\
 & & & & &  \\[-0.6cm]   {\large $|m_{e\mu}|$}  & 110 [100] MeV & 3.2 [2.5] MeV   &  0.50 [0.31] MeV & 140 [92] keV & 36 [20] keV \\ 
  \hline  & & & & & \\
 & & & & &  \\[-0.6cm]  {\large $|m_{\mu\mu}|$}  & 300 [140] MeV & 10 [3.5] MeV   &  1.5 [0.44] MeV & 420 [130] keV & 84 [28] keV \\[0.2cm]
\hline
\end{tabular}
}
\caption{\label{tab:sign-masses}
\em
The $2\sigma$ sensitivity limit on the Majorana mass matrix entries $|m_{e\mu}|$ and $|m_{\mu\mu}|$ at a muon collider. We report in brackets the sensitivity for a more optimistic scenario, where the background is reduced to a negligible level, and slightly better lepton identification efficiencies are assumed.} 
\end{table}

We apply a similar signal selection also for the process $\mu^+\mu^- \to \mu^\pm\mu^\pm W^\mp_hW^\mp_h$. In this case, we require two same-sign muons in the final state, instead of an electron and a muon and, analogously to the signal selection cuts in Eq.~\eqref{eq:cuts}, we apply
\begin{equation}\label{eq:cuts-mumu}
p_T\mu \,(1) > 2.5\%\cdot\sqrt{s}\,, \qquad p_T W_h(1) > 5\%\cdot\sqrt{s}\,,
\end{equation}
where $\mu \,(1)$ is the leading-$p_T$ muon in the final state.
The background for the $\mu\mu W_hW_h$ final state is higher than those for $e\mu W_hW_h$, due to a larger contribution from the $W_hW_h\mu^+\mu^-$ component, which is reduced by a lepton-charge misidentification factor, but it is not additionally reduced by a lepton-flavour misidentification rate. We indicate the corresponding background cross sections in Table~\ref{tab:mumu-B}, after acceptance criteria and, values in parentheses, after the cuts in Eq.~\eqref{eq:cuts-mumu}.

We can thus estimate the 2$\sigma$ sensitivities to the modulus of the effective Majorana masses $m_{e\mu}$ and $m_{\mu\mu}$, which are shown in Table \ref{tab:sign-masses}. As with the TDM analysis, we also report the sensitivities achievable in a more optimistic scenario, where the background is reduced to a negligible level, and slightly better lepton identification efficiencies are assumed.
The sensitivities of a muon collider to the Majorana neutrino masses would be significantly higher than the projected ones for the FCC-hh, in Eq.~\eqref{eq:Weinberg-hadron}, by about a factor of four for $|m_{\mu\mu}|$ and almost one order of magnitude for $|m_{e\mu}|$, already at a 3 TeV collision energy, and by more than three orders of magnitude for a 30 TeV muon collider. A similar gain in sensitivity is found with respect to a same-sign muon collider \cite{Li:2023lkl}.

Finally, let us provide a qualitative comparison between the neutrino TDM signal and the neutrino mass signal. Comparing the Feynman diagrams of Figs.~\ref{fig:NMM-muCol} and \ref{fig:wo-muCol}, one observes that, in a typical leading  diagram,
a $\lambda_{\alpha\beta}$
vertex with one derivative is traded by a $m_{\alpha\beta}$ vertex with one additional neutrino propagator and one additional electroweak coupling. Therefore, one expects roughly equal sensitivities by the replacement
\begin{equation}
\dfrac{\lambda_{\alpha\beta}}{\mu_B} 
~\leftrightarrow ~
N_{m/\lambda} \dfrac{e\cdot m_{\alpha\beta}}{\mu_B\cdot  p^2} = N_{m/\lambda} \dfrac{2m_e m_{\alpha\beta}}{p^2}\,,
\end{equation}
with $p$ the typical neutrino momentum in the process, we used $\mu_B\equiv e/(2m_e)$,
and we introduced a coefficient $N_{m/\lambda}$ to account for the different multiplicity of leading diagrams in the mass case versus the dipole case.
According to this recipe, one can compare the sensitivities in Table \ref{sensMU} or \ref{sens-CB-CW} with those in Table \ref{tab:sign-masses}. Taking into account that our simulations indicate that $N_{m/\lambda}\sim 10$,  we find that such sensitivities roughly match for a reasonable value of the neutrino momentum, $p\sim 0.05 \sqrt{s}$. This is coherent with the selection cuts in Eqs.~\eqref{eq:cuts} and \eqref{eq:cuts-mumu}.

\section{Summary and discussion 
}\label{sec:conclusions}

We presented for the first time projected collider sensitivities to the electron-muon neutrino TDM. 
We found that realistic hadron colliders cannot probe values $\lambda_{e\mu}\lesssim 10^{-9}\mu_B$. In contrast,
we showed that the LNV, $\Delta L=2$ process $\mu^+ \mu^- \to e^\pm \mu^{ \pm} W^\mp_h W^\mp_h$, at a future muon collider, could provide a much more powerful probe for the neutrino TDM.

The results of our study, in terms of sensitivities to 
$\lambda_{e\mu}$ as a function of the muon collider energy, are summarised in Table \ref{sensMU} (in Table \ref{sens-CB-CW} we give the sensitivities to the Wilson coefficients 
of the SM effective operators, which can generate such neutrino TDM).
The sensitivity strongly depends on the center-of-mass energy of the muon collider: one can cover the window between $\sim 10^{-9}\mu_B$ 
and $\sim 10^{-12}\mu_B$, as the energy grows from $\sqrt{s} \simeq  3$ TeV to 50 TeV.
In particular, we showed that a muon collider with a collision energy $\sqrt{s}\simeq 30$ TeV would reach the same level of sensitivity of the latest 
low-energy laboratory experiments.  Alternatively, the same level of sensitivity would be achieved already at $\sqrt{s}\simeq 10$ TeV, if a significant increase in the collected integrated luminosity, by a factor of the order of 100, were feasible.
Therefore, a muon collider could provide crucial complementary information on the neutrino properties, especially in the event of a near future observation at low-energy experiments.   

In order to assess the complementarity with other constraints on neutrino TDMs, we remark that, 
currently, the strongest laboratory bounds, $\lambda_\nu \lesssim 6\cdot 10^{-12} \mu_B$, comes from low-energy solar neutrinos scattering elastically on nuclei. They apply to an effective neutrino dipole moment,  therefore 
they are not directly sensitive to a specific TDM $\lambda_{\alpha\beta}$, as the flavours of the initial and final neutrino are not detected, in contrast with direct searches at colliders.
More in general, in the event of a future detection of new physics in low-energy scattering experiments, it would be difficult to identify the kind of underlying physics beyond the SM involved. For example, it would be challenging to distinguish a possible neutrino TDM signal from a dark matter scattering on nuclei. Finally, LNV cannot be established in neutrino scattering experiments. In contrast, LNV can be tested by neutrino-to-antineutrino conversion in the solar magnetic field,
however the bound is subject to the large uncertainty on the value of such magnetic field. Finally, stringent bounds on the NDMs, slightly stronger than those from low-energy neutrino scattering, 
are obtained from astrophysical and cosmological observations. These, however, might be subject to larger systematic uncertainties.

In comparison with all these other constraints, the muon collider signal would certaintly be a cleaner, more direct probe of the neutrino TDM. Actually, the muon collider could have the potential to directly access the UV physics responsible for the neutrino TDM.
This can be seen by comparing the sensitivities in Table \ref{sens-CB-CW} with the NDA estimate in Eq.~\eqref{NDA}. One observes that a positive signal could correspond to heavy particles with mass 
$m_*$ somewhat smaller than $\sqrt{s}$, which could be directly produced.
Indeed, for smaller new physics couplings to the Higgs and lepton doublets,  $g_*\epsilon_H$ and $g_*\epsilon_\alpha$, a given, observable value of $\lambda_{e\mu}$ requires a smaller $m_*$,  
therefore one may lie outside the regime of validity of the EFT. In this case the collider phenomenology might depart significantly from our analysis, in a way which strongly depends on the specific UV completion.
In contrast, for larger couplings, the heavy states could lie above the muon collider energy and the
effective operators remain a good approximation. In this second case, though, a specific model of lepton flavour is required, to avoid that the same couplings violate bounds on $e-\mu$ transitions in low-energy experiments.
Both scenarios appear very interesting to explore, and require non-trivial model building, which is left for future work.

We also studied the same process, $\mu^+\mu^-\to \ell^\pm_\alpha \ell^{\pm}_\beta W^\mp_h W^\mp_h$, when the source of LNV is not a NDM, but rather a Majorana neutrino mass, $m_{\alpha\beta}$. We estimated projected  sensitivities on $|m_{\mu\mu}|$ and $|m_{e\mu}|$, shown in Table \ref{tab:sign-masses}. At 2$\sigma$, the sensitivity to $m_{e\mu}$ ranges from 
$\sim 100$ MeV at a 3 TeV muon collider to $\sim 30$ keV at $\sqrt{s} \simeq 50$ TeV, with slightly weaker sensitivities in the case of $m_{\mu\mu}$. This would represent an improvement of up to three orders of magnitude, with respect to the sensitivities of the FCC-hh, which have been estimated recently in \cite{Fuks:2020zbm}. To the best of our knowledge, these would be by far the strongest direct bounds on these neutrino-mass-matrix elements.

What if a large same-sign dilepton signal is observed at a muon collider, incompatible with current bounds on Majorana neutrino TDMs and masses?
Indeed, $|\lambda_{e\mu}|\gg 10^{-11}\mu_B$ would be very difficult to reconcile with the various complementary constraints. Similarly, $|m_{\alpha\beta}|\gg 1$ eV
would be incompatible with the minimal Majorana-neutrino-mass scenario, 
given the upper bound $|m_{ee}|\lesssim 0.1$ eV from $0\nu2\beta$-decay searches.
In these cases, a positive signal at the muon collider would point to additional, LNV new physics.
This may correspond either to new light states, e.g.~sterile neutrinos with masses below the collider energy, or to additional LNV higher-dimensional operators, as discussed at the end of section \ref{sec:theo}.

Alternatively, in the case of a positive signal in a low-energy experiment, compatible with a NDM $\lambda_\nu\gtrsim 10^{-12}\mu_B$, we have shown that a muon collider would have the capability to confirm or disprove the interpretation of such signal in terms of a Majorana neutrino TDM,
by a direct observation of lepton number and flavour violation.

\acknowledgments

We thank Marco Ardu, Sacha Davidson, Sudip Jana, Danny Marfatia and Daniele Montanino, for conversations on neutrino dipole moments, and Richard Ruiz, for discussions of the event simulations. The authors would like to express special thanks to the Mainz Institute for Theoretical Physics (MITP) of the Cluster of Excellence PRISMA+ (Project ID 39083149), for its hospitality and support during the MITP Capri Program 2022 on `Neutrinos, Flavour and Beyond'.
MF has received support from the European Union Horizon 2020 research and innovation program under the Marie Sk\l odowska-Curie grant agreements No 860881-HIDDeN and No 101086085–ASYMMETRY.
The work of NV is supported by ICSC – Centro Nazionale di Ricerca in High Performance Computing, Big Data and Quantum Computing, funded by European Union – NextGenerationEU, reference code CN\_00000013.


\end{document}